# Understanding and Design of Interstitial Oxygen Conductors

Jun Meng

*Department of Materials Science and Engineering, University of Wisconsin Madison, Madison, WI, USA.*

Highly efficient oxygen-active materials that react with, absorb, and transport oxygen is essential for fuel cells, electrolyzers and related applications. While vacancy-mediated oxygen-ion conductors have long been the focus of research, they are limited by high migration barriers at intermediate temperatures (400–600°C), which hinder their practical applications. In contrast, interstitial oxygen conductors exhibit significantly lower migration barriers enabling faster ionic conductivity at lower temperatures. This review systematically examines both well-established and recently identified families of interstitial oxygen-ion conductors, focusing on how their unique structural motifs such as corner-sharing polyhedral frameworks, isolated polyhedral, and cage-like architectures, facilitate low migration barriers through interstitial and/or interstitialcy diffusion mechanisms. A central discussion of this review focuses on the evolution of design strategies, from targeted donor doping, element screening, and physical-intuition descriptor material screening and machine learning approach, which leverage computational tools to explore vast chemical spaces in search of new interstitial conductors. The success of these strategies demonstrates that a significant, largely unexplored space remains for discovering high-performing interstitial oxygen conductors. Crucial features enabling high-performance interstitial oxygen diffusion include the availability of electrons for oxygen reduction and sufficient structural flexibility with accessible volume for interstitial accommodation and migration. This review concludes with a forward-looking perspective, proposing a knowledge-driven methodology that integrates current understanding with data-centric approaches to identify promising interstitial oxygen conductors outside traditional search paradigms. These approaches are expected to significantly accelerate the development of high-performance interstitial oxygen conductors for a variety of oxygen-active applications, ultimately paving the way for more efficient and sustainable energy technologies.

## Contents

# 1. Introduction

Oxygen-active materials play a crucial role in high-efficient energy conversion applications for a sustainable future, such as solid oxide fuel cells (SOFCs), [1–4] electrolyzers, [5] solid-oxide metal-air redox batteries, [6–9] gas sensors, [10,11] gas separation membranes [12] for fuel production and gas purification, chemical looping,[13] anomia synthesis[14] and more. For instance, SOFCs can operate with diverse fuel sources such as $H_2$, CO, $CH_4$, and various hydrocarbon blends. The electrochemical oxidation process in SOFCs inherently maintains high energy conversion efficiency and low pollutant emissions. However, the widespread adoption of SOFC technology requires the development of fast oxygen-ion conducting materials that exhibit both high ionic conductivity and robust physicochemical stability at intermediate temperatures (400–600˚C). The current commercial oxygen-ion conductors rely on vacancy-mediated oxygen transport, where oxygen vacancies serve as mobile charge carriers. Notable examples include fluorite-type yttrium-stabilized zirconia (YSZ), cerium gadolinium oxide (CGO), and perovskite-type strontium- and magnesium-doped lanthanum gallate $La_{0.9}Sr_{0.1}Ga_{0.8}Mg_{0.2}O_{3-\delta}$ (LSGM). However, these materials only have adequate oxygen kinetics for practical applications at high temperatures $\approx 800$ ˚C, primarily due to their relatively high oxygen-ion migration barriers (>1 eV). Such high operating temperatures increases cost and accelerate materials degradation, impeding their use for broader implementation and long-term viability.

Efforts to discover and develop oxygen-ion conductors with lower migration barriers are crucial for achieving robust ionic conductivity at intermediate temperatures. At the atomic level, oxygen-ion migration typically follows one of these main diffusion mechanisms: direct lattice diffusion, vacancy diffusion, interstitial diffusion, and interstitialcy diffusion, depending on the material's crystal structure and the presence of various defects, such as oxygen vacancies, interstitials, Frenkel pairs, and Schottky defect. In direct lattice diffusion, oxygen-ion diffuses directly through lattice without defects. Although in some cases, oxygen self-diffusion directly through lattice exhibiting non-travail diffusivity, such as in $UO_2$,[15] it is generally orders of magnitude slower than defect-assisted diffusion and is not a primary pathway in most practical materials. In vacancy-mediated diffusion, oxygen ions diffuse by hopping into adjacent oxygen vacancies in a partially occupied oxygen sublattice. This mechanism generally demands high migration barriers (>1 eV), as each hop involves bond breaking and re-forming. For efficient conduction, a uniform set of energetically equivalent lattice sites, such as the anion sites in a fluorite structure, is considered ideal for supporting vacancy-mediated oxygen-ion transport.[16] Ever since early 20th-century studies on oxygen diffusion in fluorite-based materials, researchers have primarily aimed to enhance ionic conductivity by examining structurally analogous oxides and introducing oxygen vacancies. This approach has guided the development of many high-performing oxygen-ion conductors: such as (1) fluorite-related oxides like YSZ, CGO,[17] $Bi_2O_3$,[18] and pyrochlore-type oxides;[19] (2) perovskite-type materials LSGM,[20,21] Brownmillerite,[22] Ruddlesden-Popper,[23] and layered perovskite;[24] (3) Aurivillius-type materials[25–27] and BIMEVOX series[28] consisting of fluorite-like and perovskite-like layers, (4) LAMOX families,[29] and more. Interstitial oxygen occupies interstitial sites rather than existing lattice positions. In the direct interstitial mechanism, interstitial oxygen moves from one interstitial site to another, bypassing the regular lattice. For the interstitialcy (or "cooperative") mechanism, an interstitial oxygen ion "kicks out" a lattice oxygen atom from its site, pushing it into another interstitial position, which incorporates both interstitial diffusion and host-atom displacement. This dynamic process leads to rapid, collective defect migration and enables high oxygen conductivity. Interstitial and interstitialcy diffusion are typically more rapid than vacancy diffusion, as the diffusing species encounters less resistance in the form of lattice distortions or strain. High-performing interstitial oxide ion conductors are relatively rare, compared with conventional oxygen vacancy conductors. These materials are usually less densely packed, which could be favorable for accommodating extra oxygen atoms, and the polyhedral could facilely deform, offering room for forming and breaking chemical bond with the interstitial oxygen. Oxides presently known to have predominantly interstitial-mediated oxygen conduction span various structural families including fluorite,[30,31] sillén oxychlorides,[32–34] scheelite,[35,36] apatite,[37] hexagonal perovskite derivatives and palmierite,[38–40] Ruddlesden-Popper pervoskite,[41] melilite,[42] hexagonal manganite,[43] mayenite,[44] langasite,[45] garnet,[46,47] and perrierite.[48]

The temperature (T) dependence of the diffusion coefficient (D) has an Arrhenius form of $D = D_0 exp\left(-\frac{E_a}{k_B T}\right)$, where $D_0$ is the prefactor and $E_a$ is the activation energy of diffusion, i.e. the sum of the migration energy $E_m$ and the formation energy $E_f$ of the diffusing defects. The activation energy required for oxygen-ion migration is a key factor in determining the diffusion coefficient and, consequently, the ionic conductivity. Lower activation energies lead to higher ionic conductivity at a given temperature. In terms of this, interstitial oxygen-ion conductors have many potentially advantages over vacancy-mediated conductors yet received limited research attentions until very recently. First, interstitial oxygen conductors typically have lower migration barriers compared to vacancy oxygen conductors. This observation is supported by the experimental migration barriers for oxygen ion-conducting oxides. In **Figure 1**, data collection from the Citrine Informatics ARPA-E Ionics Database,[49] as well as recent experimental studies shown that interstitial conductors have an average migration barrier of about 0.6 eV, compared to around 1.0 eV for vacancy-mediated systems.[48] This 0.5 eV reduction can translate into a roughly 1000 times increase in ionic conductivity at 600 °C. Additionally, under a given oxygen partial pressure, interstitial oxygen tends to be more thermodynamically stable at lower temperatures, enabling higher defect concentrations and thus higher conductivity, opposite the trend for vacancy-based materials. Moreover, oxygen surface exchange and catalytic reactions involving pulling oxygen to the surface are potentially faster when transport is mediated by interstitials as compared to vacancies, as the entire surface has accessible interstitial sites to absorb oxygen. The synergistic combination of lower migration barriers, increasing concentration at low temperatures, and many active surface sites for exchange suggest that interstitial oxygen conductors may lead to great performance improvements in oxygen-active materials.

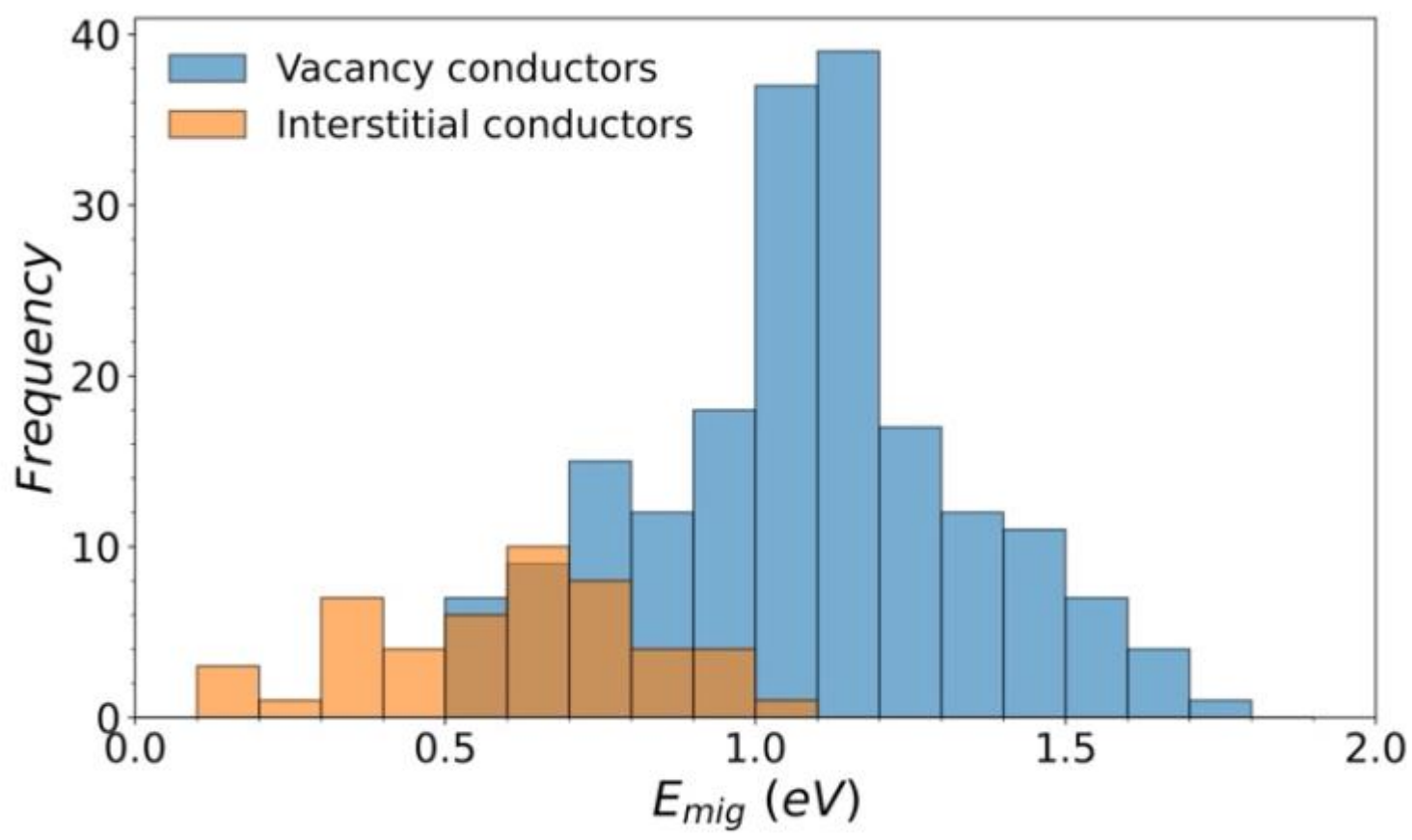

Figure 1 Migration barrier comparison of known interstitial and vacancy diffuser oxides obtained from the Citrine Informatics database and literature. Figure adapted with permission from Ref. [48].

Therefore, a deeper understanding of interstitial oxygen conduction mechanisms is essential for designing next-generation fast oxygen-ion conducting materials, particularly for intermediate temperature applications. This knowledge is pivotal in addressing the current limitations of existing materials and advancing discovery and design of new high-performing ones. To the best of our knowledge, no comprehensive review has yet been conducted that systematically summarizes the diverse families of interstitial oxygen conductors. In this review, we fill this gap by offering an in-depth analysis of these material families, focusing on their structural and chemical features, performance metrics, and the strategies employed for their discovery and design. Through a comprehensive review on these known interstitial oxygen conductors, it is summarized that crucial chemical and structural features governing the ability of a material to effectively incorporate and diffuse interstitial oxygen-ion are the availability of electrons for oxygen reduction and the structural flexibility with sufficient accessible volume for facile oxygen diffusion. Leveraging these insights, this review proposed a forward looking, knowledge-driven approach to identify new promising interstitial oxygen conductors, distinct from traditional material searches, aiming to accelerate the advancement of oxygen-active materials related applications.

## 2. Interstitial Oxygen Conducting Families

In this section, we highlight the discovery and evolving understanding of representative material families of interstitial oxygen conductors. For each family, we review their crystal structural features, ionic conductivity performance, and the underlying structure–property relationships that enable or limit interstitial oxygen transport. This approach provides a comprehensive view of diverse structural motifs govern the formation and migration of interstitial oxygen ions. Our review of these known families reveals two consistently critical features that are essential for the incorporation of interstitial oxygen ions and for promoting their rapid diffusion: (1) the availability of electrons, typically supplied by oxidizable cations or aliovalent doping, and (2) structural flexibility with sufficient free volume, often provided by corner-sharing polyhedral networks, isolated polyhedra and their combinations, or highly symmetric anion framework with well-defined interstitial sites such as fluorite.

To guide this review, we organize the known interstitial oxygen conductors according to their key structural features critical for interstitial incorporation and diffusion: (i) Fluorite $UO_{2+x}$ and sillén oxychloride, which possess highly symmetric oxygen sublattices with well-defined interstitial sites facilitating fast interstitialcy diffusion; (ii) Scheelite, apatite, hexagonal perovskite derivatives and palmierite, characterized by isolated tetrahedra that provide local structural flexibility and open pathways for ion migration; (iii) Ruddlesden-Popper perovskite, melilite, and hexagonal manganite, exhibiting two-dimensional corner-sharing polyhedral networks that accommodate planar diffusion pathways; (iv) Mayenite, langasite, and garnet, featuring three-dimensional corner-sharing polyhedral frameworks offering interconnected diffusion channels; and (v) Perrierite/chevkinite $La_4Mn_5Si_4O_{22+\delta}$, representing a combination of isolated sorosilicate units and corner-sharing networks which enabled both interstitial and interstitialcy diffusing mechanisms.

### 2.1 Fluorite $UO_{2+x}$

$UO_2$, commonly used as nuclear fuel, crystallizes in a fluorite structure where the oxygen sublattice can accommodate excess oxygen in interstitial sites, which play a significant role of the oxygen self-diffusion and micro-structure evolution of $UO_2$. $UO_2$ is known to exhibit varying oxygen contents depending on the temperature and oxygen partial pressure. As excess oxygen is introduced, oxygen interstitials form in the lattice as point defects at low x in $UO_{2+x}$. As the oxygen content x increases, interstitials oxygen defects aggregate to form clusters,[50,51] and even lead to the transition to hyperstoichiometric phases such as $U_4O_9$. Here we focus on simplest scenario about the formation and diffusion of interstitial oxygen as point defects in the $UO_2$ lattice. As shown in **Figure 2a**, $UO_2$ maintains a fluorite structure where U ions (cyan spheres) occupy the 4a sites and O ions (red spheres) occupy the 8c sites in a face-centered cubic arrangement. Interstitial oxygen occupies the octahedral site (black spheres), which is the center of the fluorite unit. *Ab initio* studies revealed that the interstitial oxygen migrate via an interstitialcy mechanism, where the interstitial oxygen atom displaces one lattice oxygen and push it to another interstitial site, with the migration barriers of 0.81–0.93 eV.[31] P. Garcia *et al.* combined electrical conductivity measurements with $^{18}O$ isotopic exchange to analyze oxygen diffusion properties of $UO_2$,[52] demonstrated that in the intrinsic regime, oxygen diffusion is mediated by doubly charged interstitial oxygen, corresponding to an increase in the concentration of oxygen interstitials as $U^{4+}$ ions were oxidized to $U^{5+}$ states. **Figure 2b**, displayed the measured conductivity of the synthesized single crystal, polycrystal $UO_2$, and Cr-doped polycrystalline $UO_2$ in the extrinsic regime, in which the conductivity of single crystal $UO_2$ reaches up to 1.1 S/cm at 1000˚C.[31,52]

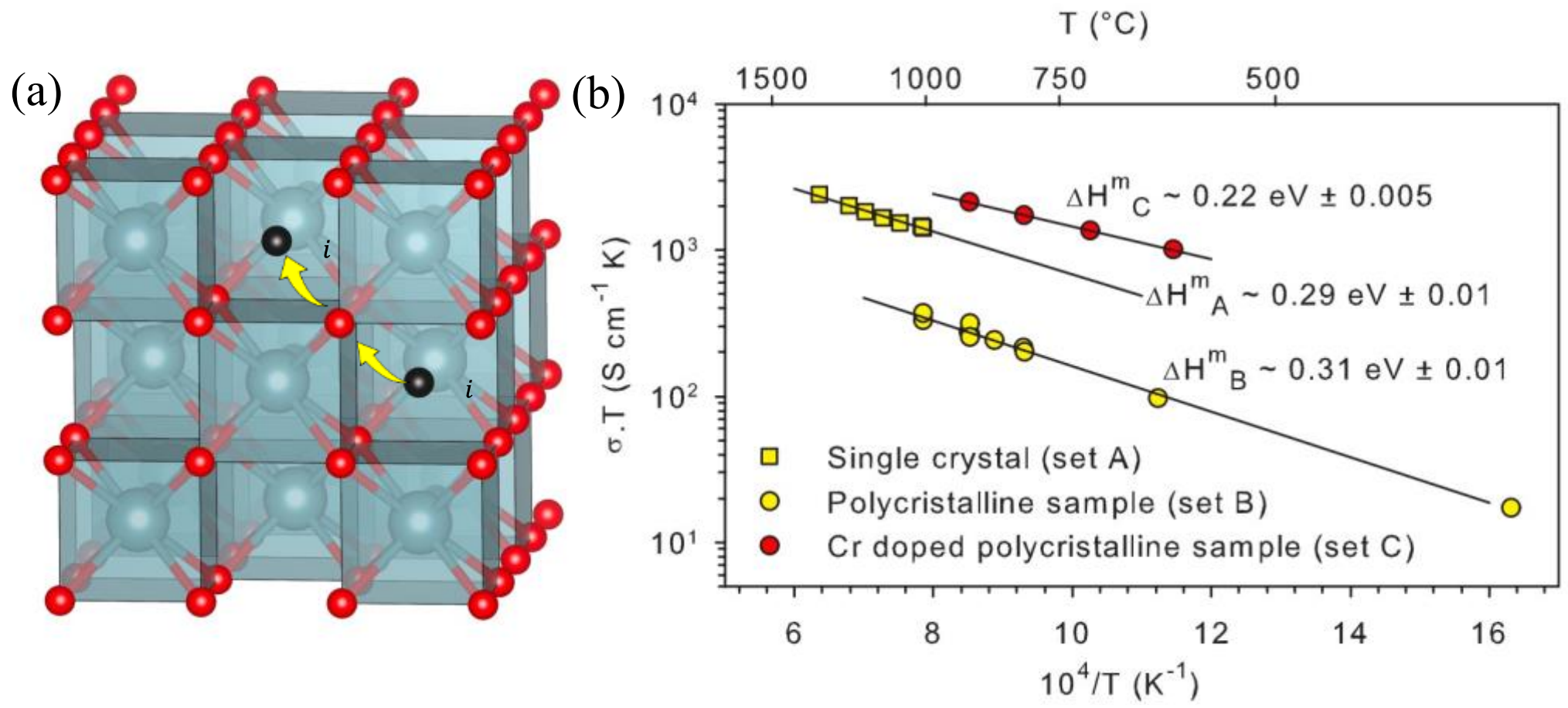

Figure 2 (a) Structure of $UO_2$. The $U^{4+}$, $O^{2-}$ atoms and $O_i$ sites are shown as cyan, red, and black spheres, respectively. The yellow arrows indicate the diffusion pathways via the interstitialcy mechanism. (b) Electrical conductivity of single crystal and polycrystalline $UO_2$, and Cr-doped polycrystalline $UO_2$ in the extrinsic conductivity regime. Figure adapted with permission from Ref. [31]

It is well-known that fluorite-type structures can accommodate both interstitials and vacancies, and many fluorite materials oxygen deficiency, such as YSZ, CGO, $Bi_2O_3$, and pyrochlore, are exceptional oxygen-ion conductors. Notably, $UO_2$ stands out in this family due to its ample space within the lattice and the facile valence change of $U^{4+}$ to $U^{5+}$, making it capable of hosting and diffusing excess oxygen ions.[53] Although $UO_2$ is not widely used in oxygen-ion conducting applications due to its radiative properties, its unique structural and electronic characteristics make it an representative interstitial oxygen-ion conducting material for further study.

### 2.2 Sillén oxychloride

Bismuth-based Sillén oxychlorides, with the general formula $MBi_2O_4X$ (M=rare-earth element, X=halogen element), have recently been recognized as interstitial oxygen-ion conducting materials since 2023.[32–34] Sillén oxychlorides features a triple fluorite-like layer structure, where layers of $Bi_2O_2$ are sandwiched between $MO_2$ cubes, shown in **Figure 3a**. This "triple fluorite" framework is essential for hosting defects and enabling efficient ion conduction. High oxygen-ion conduction in Sillén oxychlorides with excess oxygen is uncovered and attributed to interstitial oxygen-ion conduction via the interstitialcy diffusion mechanism.

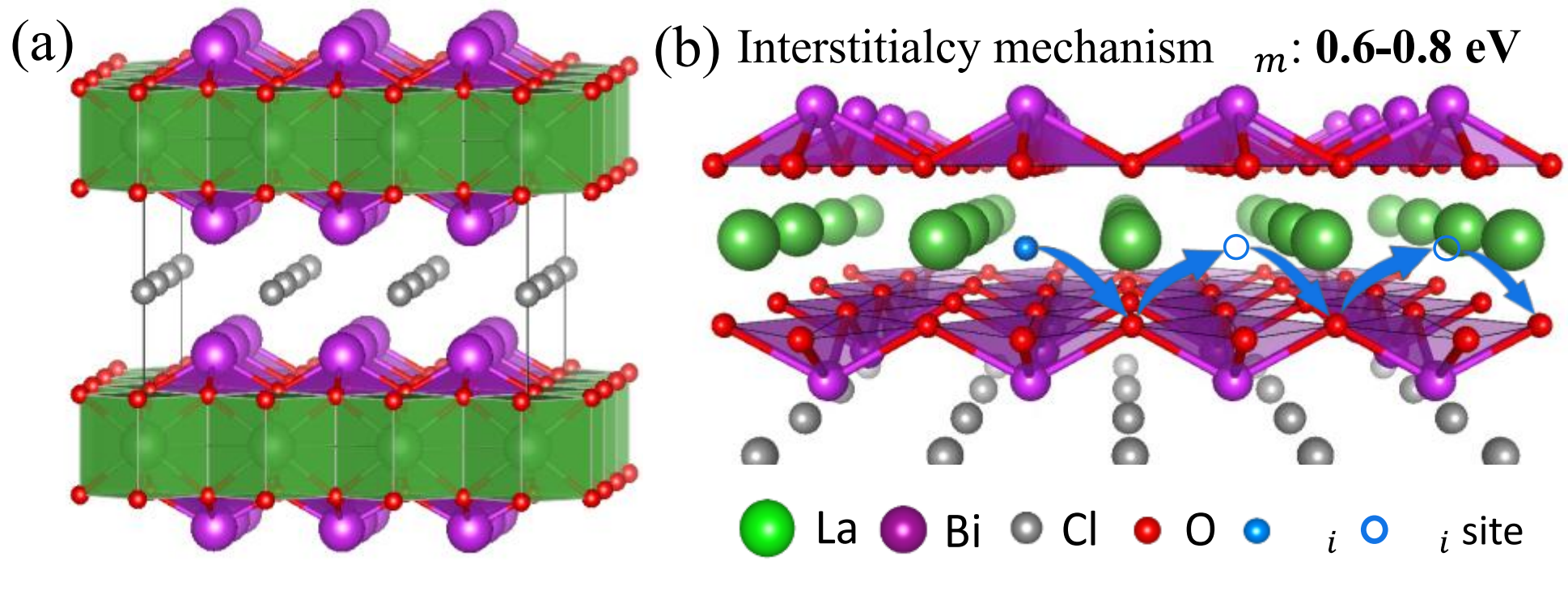

Figure 3 (a) Structure of $LaBi_2O_4Cl$ (LBC). The $La^{3+}$, $Bi^{3+}$, $Cl^-$ and $O^{2-}$ atoms are shown as green, purple, grey, and red spheres, respectively. (b) Illustrations of the diffusion mechanism of interstitial $O_i''$. The migration barriers are calculated by the climbing image nudged elastic band (CI-NEB) method. Figures adapted with permission from Ref. [32].

Yaguchi *et al*. demonstrated high oxygen-ion conductivity in Sillén oxychlorides $LaBi_{1.9}Te_{0.1}O_{4.05}Cl$ with a high conductivity of 0.02 S/cm at 702 °C.[33] In addition, $LaBi_{1.9}Te_{0.1}O_{4.05}Cl$ exhibited and displayed excellent chemical and electrical stability under a wide range of conditions, including various oxygen partial pressures ($P(O_2)$), $CO_2$ flow, reductive wet 5% $H_2$ in $N_2$ flow, and static air with natural humidity at 400 °C. High-temperature neutron diffraction, crystal structure analysis, maximum entropy method (MEM) neutron scattering length density (NSLD) analysis, bond valence energy (BVE) calculations, and *ab initio* calculations together have revealed that the observed high ionic conductivity is due to the presence of interstitial oxygen created by Te-doping. Building on this finding, Ueno *et al*. expanded the composition to the $LnBi_{1.9}Te_{0.1}O_{4.05}Cl$ series (Ln = Nd, Sm, Eu, Gd, Dy, Ho, Er, Tm, Yb, Lu) and identified that $LuBi_{1.9}Te_{0.1}O_{4.05}Cl$ exhibits a high bulk conductivity of 0.022 S/cm at 492°C, higher than commercial material YSZ.[34]

Meanwhile, Meng *et al.* performed a structure-similarity analysis of ≈ 62k oxygen-containing compounds using fluorite as a reference structure, and identified $LaBi_2O_4Cl$ (LBC) as a fast oxygen-ion conductor enabled by both oxygen vacancy and interstitial diffusion.[32,54] *Ab initio* studies on LBC have elucidated the defect chemistry that the Frenkel pair defect is energetically most favored. Despite the formation of interstitial oxygen or oxygen vacancies is challenging in intrinsic LBC, doping with higher ($Te^{4+}$) or lower ($Sr^{2+}$) valence cations have been shown to activate ion conduction.[32,33] The "triple fluorite" structure accommodates both oxygen vacancies and interstitial oxygen ions, supporting both vacancy-mediated and interstitial-mediated diffusion two-dimensional in the "triple fluorite" layer, according to *ab initio* studies. NEB calculations indicated that he migration barrier of vacancy and interstitial diffusion is 0.1 eV and 0.6 eV, respectively. These interstitial oxygen-ions locate at the cubic voids of the FCC $La^{3+}$ lattice, and migrate through the interstitialcy mechanism within the "triple fluorite" layers, as shown in **Figure 3b**.

These findings highlighted the broader potential of fluorite-derived structures, especially those with layered or distorted fluorite motifs such as sillén oxychlorides, as promising candidates for interstitial oxygen-ion conduction. Their structural frameworks with cubic voids and highly symmetric oxygen sublattices, provide well-defined interstitial sites and diffusion pathways. Moreover, the ability to support both vacancy and interstitial transport, coupled with exceptional chemical and electrical stability, positions these materials as strong candidates for next-generation low-temperature solid electrolytes. Future advancements through targeted doping and structural engineering are likely to further enhance their performance.

### 2.3 Scheelite

Scheelites, with general formula $ABO_4$, exist in a wide range of compositions and have a rather open structure (**Figure 4a**) that can be thought of as a modification of the fluorite structure with an ordered arrangement of two cations A and B. The larger cations A show eight-fold coordination, and the smaller cations B are four-fold coordinated.[55] Early evidence of oxygen ion transport in scheelite dates back to studies on $PbMoO_4$ and $PbWO_4$ in 1970-1990s. Researchers revealed that $PbMoO_4$ crystals exhibited isotropic pure ionic conduction below ~360 °C,[56] indicating that defects in the oxygen sublattice contributes significantly to the conductivity in this family, most likely due to oxygen vacancies as $PbMoO_4$ crystals grown with some loss of oxygen. T. Lu and B.C.H. Steele performed impedance spectroscopy measurement on the electrical conductivity of $BiVO_4$ and found that the ionic conductivity is due to oxygen-ion vacancies with an estimated relatively low migration barrier of ~ 0.33 eV in the low temperature regime (140-300 °C).[57]

In addition to the earlier recognition of vacancy-mediated diffusion, scheelite has also been identified as an interstitial oxygen conductor. The disparity in cation sizes induces distortion in the anion lattice, creating accessible space for interstitial oxygen incorporation and migration. Over the past few decades, neutron diffraction, conductivity measurements, and computational investigations have solidified the concept of interstitial oxygen-ion conduction in various scheelite materials. Early efforts to induce oxygen excess and thus promote potential interstitial conduction in scheelites commonly employed aliovalent donor doping. Doping with lower valent cations can generate oxygen vacancies, whereas doping with higher valent cations generates interstitial oxygen. Cava *et al.* studied excess-oxygen in Th-doped $LaNbO_4$ ($La_{0.93}Th_{0.07}NbO_{4.035}$) and observed significantly higher ionic conductivity, approximately 3 orders of magnitude higher compared to $LaNbO_4$.[58] In addition, experiments confirmed oxygen interstitials in W-doped $LaNbO_4$ ($LaNb_{1-x}W_xO_{4+x/2}$) and improved conductivity with doping levels of x ~ 0.08-0.16, with $LaNb_{0.92}W_{0.08}O_{4.04}$ exhibiting roughly 0.1 S/cm at 1000 °C.[59] Hoffart *et al.* uncovered that $BiVO_4$ exhibits a combination of ionic and electronic conductivity that can be strategically tuned through doping.[60] In this study, $BiVO_4$ was found to support oxygen-ion diffusion through either vacancy or interstitial mechanism, depending on lower or higher valent cation substitution. Notably, Mo-doped $BiVO_4$, corresponding to interstitial oxygen diffusion, shown a total conductivity approximately two orders of magnitude higher than that of YSZ. Furthermore, high-valent cations B can stabilize the $BO_4$ tetrahedra whereas high-valent cations A may destabilize the rigid $BO_4$ coordination, reducing the strength of the B-O bonds and increasing the oxygen mobility. This insight suggests that high-valent doping at the A-site may promote interstitial oxygen mobility and thereby increase ionic conductivity. A compelling example is the substitution of $Pb^{2+}$ with $La^{3+}$ in $PbWO_4$, which introduces excess oxygen to maintain charge neutrality. Experimental studies have shown that $La_{0.2}Pb_{0.8}WO_{4.1}$ exhibits an oxygen ionic conductivity of 0.042 S/cm at 800 °C, with an activation energy of 0.59 eV in the high-temperature regime.[61] In another case, $CeNbO_{4+\delta}$ demonstrated that the mixed-valence nature of $Ce^{3+}/Ce^{4+}$ facilitates the accommodation of oxygen interstitials, leading to a conductivity of approximately 0.03 S/cm at 850 °C.[35,36,62]

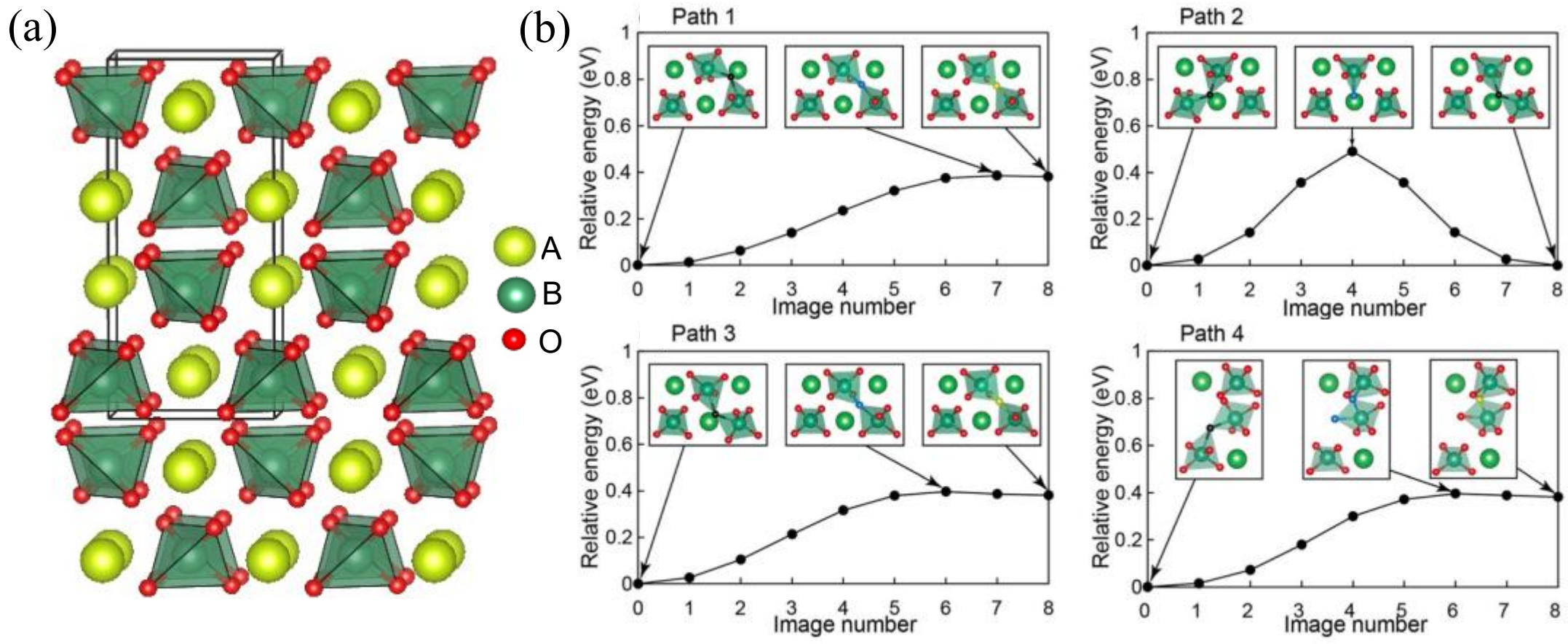

Figure 4 (a) Atomic configuration of Scheelite $ABO_4$. (b) The energy profiles along diffusion paths 1–4 observed in scheelite $LaNbO_4$. The local structures of the initial, final, and saddle-point states are also inserted in each figure. The black, yellow, and red spheres denote oxygen ions at the Oint-1, Oint-2, and regular sites, respectively. The blue spheres are oxygen ions migrating at the saddle-point state in each path, which occupy different sites between the initial and final states. Figures adapted with permission from Ref. [63].

*Ab initio* and classical MD studies demonstrate that interstitial oxygen in scheelite undergoes migration through two main mechanisms: (i) direct hops between neighboring interstitial sites and (ii) a knock-on (interstitialcy) process. As shown in **Figure 4b**, DFT calculations identify two particularly stable interstitial sites in $LaNbO_4$ with low migration barriers along the resulting pathways. In Path 1–3, the additional oxygen hops directly between these identified interstitial sites, whereas Path 4 follows the interstitialcy mechanism where an interstitial oxygen displaces a lattice oxygen, creating a new interstitial configuration as it migrates. This interstitialcy mechanism was found prevalent in La-doped $PbWO_4$, where the neutron diffraction and X-ray absorption fine structure (XAFS) studies found that part of these extra oxygen ions reside near the $WO_4$ tetrahedra, forming $W_2O_9$-like clusters.[64] MD simulations shown that the interstitial oxygen-ion can move among the $WO_4$ tetrahedral units through a synergic mechanism involving continuous breaking and reformation of $W_2O_9$ units assisted by rotation and deformation of $WO_n$ polyhedra and knock-on process between oxygen atoms.

Studies on scheelites accommodating and transporting interstitial oxygen ions have shown that their loosely packed isolated tetrahedral sublattice offer ample space and tolerate a wide range of dopants on both A and B sites, leading to the formation of stable interstitial oxygen defects and enabling fast diffusion via interstitial and interstitialcy mechanisms. A key feature in scheelite family is the structural flexibility of the $BO_4$ tetrahedra, their ability to readily rotate, deform, and transiently connect via corner- or edge-sharing configurations plays a critical role in both stabilizing interstitial oxygen and facilitating its migration through the lattice.

### 2.4 Apatite

The study of silicate apatites as fast oxygen ion conductors began in 1995 with the work of Nakayama *et al.*,[65,66] who made notable contributions demonstrating lanthanoid silicates $Ln_x(SiO_4)_6O_{1.5x-12}$ (Ln=La, Nd, Sm, Gd, Dy, Y, Ho, Er, and Yb, x=8-11), featuring a hexagonal apatite structure (space group: P63/m) with a range of x from 8 to 9.33, were found to exhibit ionic conductivity of 4.3 to $7.9\times10^{-3}$ S/cm at 300 °C, comparable to YSZ and LSGM.[66] Structure of silicate apatite consists of isolated $SiO_4$ tetrahedra and accommodates a large variety of cations occupying channels running along the c-axis, as shown in **Figure 5a**. In this apatite structure, oxide ions which are not a member of the $SiO_4$ tetrahedron are located at the O4 site of the channel along the c-axis, are responsible for the ionic conduction. Neutron diffraction investigations have revealed the precise atomic structures of silicate apatite such as $La_{9.55}Si_6O_{26.32}$, $La_{9.33}Si_6O_{26}$ and $La_8Si_6O_{26}$, providing insight for understanding the ionic conduction mechanism. In $La_{9.55}Si_6O_{26.32}$, the interstitial oxygen-ion on O5 site next to La atoms and a $SiO_4$ tetrahedron was revealed and considered responsible for oxygen-ion conduction, shown as orange spheres in **Figure 5b**. In $La_{9.33}Si_6O_{26}$, the O4 sites disorder was found to enhance ionic conductivity compared to $La_8Si_6O_{26}$.[67,68]

Subsequent computational modeling clarified the defect chemistry and oxygen-ion diffusion mechanisms in silicate apatite,[69–78] demonstrating that oxygen-ion diffusion is mediated by both interstitial and interstitialcy mechanism. Tolchard *et al.* performed static lattice simulations of $La_{9.33}Si_6O_{26}$ and proposed that interstitial oxygen-ions on O5 sites present and diffuse through a non-linear "sinusoidal-like" pathway along the c axis.[69] Kendrick *et al.* performed MD simulation of $La_{9.33}Si_6O_{26}$, supported the sinusoidal-like interstitial mechanism in the O4 channels, and uncovered that the interstitial oxygen-ions also diffuse normal to the c axis via interacting with $SiO_4$ units.[75] Béchade *et al.* applied static lattice simulations and the bond valence method for stoichiometric $La_{9.33}Si_6O_{26}$ and proposed oxygen-ion diffuses via an indirect-interstitial mechanism. In this process, channel oxygen-ions on O4 sites are involved in the conduction via a push-pull mechanism with a calculated migration energy of 0.32 eV.[78]

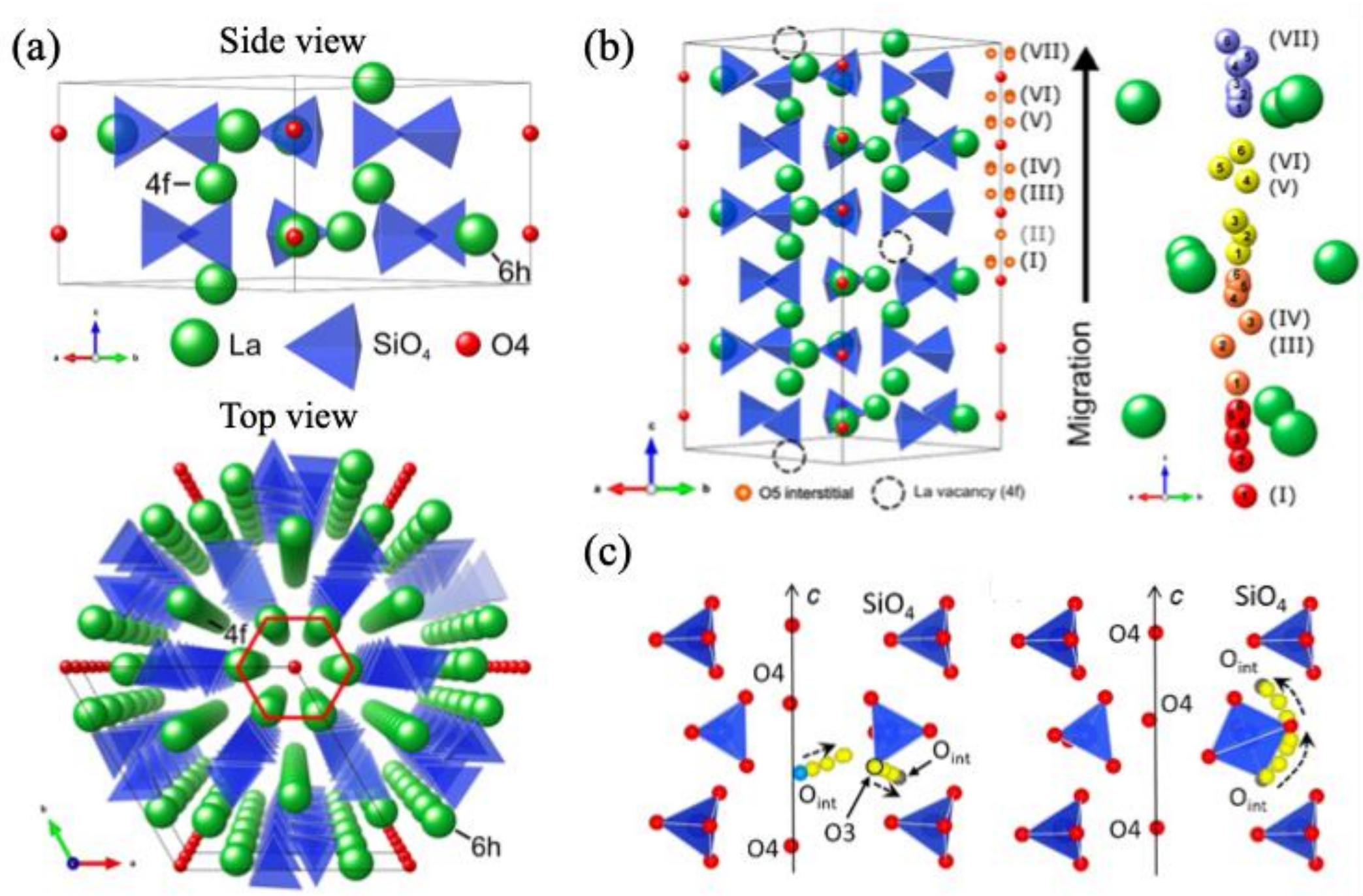

Figure 5 (a) Side view (up) and top view (down) of the structure of lanthanum silicate apatite. The La 4f sites in the center of the unit cell form pillars in c-axis direction, whereas the La 6h sites form a triangular coordination of O4 oxygen ions. The hexagonal La-tunnel is highlighted in red with the mobile O4 oxygen ions inside. (b) Left: 1 × 1 × 3 supercell of $La_{9.33}Si_6O_{26}$ orthogonal to the c-axis. The stable oxygen interstitial sites (O5), arranged in a layer structure with three equivalent positions per layer, are displayed in orange. This La vacancy distribution leads to six different O5 interstitial layers. Right: Diffusion pathways of interstitial oxygen ions at the O5 site along the c-axis, occurring via an interstitialcy mechanism, where the interstitial oxygen on O5 pushes an oxygen-ion on O4 into another interstitial site. Each color represents a distinct oxygen-ion. (c) Left: Diffusion pathways of interstitial oxygen-ions in the a-b plane, where an interstitial oxygen-ion hops from the O5 site to the metastable interstitial site between two $SiO_4$ tetrahedra. Right: Diffusion pathways of interstitial oxygen-ions that hop between $SiO_4$ tetrahedra along the c axis via the interstitialcy mechanism. Figures adapted with permission from Ref.[71,72].

A variety of atomic-level diffusion pathways for ionic conduction have been proposed. Here, we highlight the most prevalent diffusion pathways with lowest migration barriers in lanthanum silicate apatite, based on DFT studies. First, the most stable and metastable interstitial oxygen sites, as determined by DFT, are located at the O5 sites near the O4 columns, and between two neighboring $SiO_4$ tetrahedra (see **Figure 5c**), respectively.[72] The interstitial oxygen-ion diffuses along the c-axis via an interstitialcy (push-pull, or coorperative) mechanism, where the interstitial oxygen-ion on O5 site pushes an O4 oxygen-ion into another interstitial site. This mechanism was found feature a lower migration barrier of 0.3 eV by nudged elastic band (NEB) calculation.[71] In **Figure 5c**, the left panel displayed the interstitial oxygen-ion hops from the O5 site to the metastable interstitial site between two $SiO_4$ tetrahedra involving sequential bond-forming and bond-breaking events of Si-O during the conduction, with a migration barrier of 0.58 eV.[72] The right panel shows the interstitial oxygen-ion diffusing between $SiO_4$ tetrahedra, facilitated by the rotation of the tetrahedra, featuring an energy barrier of 0.38 eV.[79] Due to this channel features and the relatively lower migration barriers along c-axis, lanthanum silicate apatite shown significant anisotropic ion conduction, observed in many experimental studies.[80,81]

In germanate apatite $La_xGe_6O_{12+1.5x}$ (x=8–9.33), research revealed isotropic and significantly higher oxygen-ion conduction compared to silicate apatite, reaching to approximately 0.01-0.08 S/cm at 800 °C.[82–86] Atomistic simulations shown that the most stable interstitial oxygen sites located at the center of two $GeO_4$ tetrahedra forming the ‘‘$Ge_2O_9$’’ units, as shown in **Figure 6**. In contrast, *ab initio* studies indicated that there are no low-energy interstitial sites in the O4 channel.[72] On the experimental side, neutron diffraction analyses revealed interstitial sites near the $GeO_4$ and the formation of $GeO_5$ units, as well as at the periphery of the O4 channel. These results indicated that the interstitial oxygen in germanate apatite located either at the channel or between two $GeO_4$ tetrahedra, with the latter likely being more prevalent. Atomistic simulation and MD simulations[76,86–88] revealed that the channel O4 oxygen ions act as a reservoir for the formation of oxygen interstitials, which then migrates either along the c-axis or in the ab-plane by forming and breaking of ‘‘$Ge_2O_9$’’ units, with migration barriers of 0.64 eV and 0.76 eV, respectively, as shown in **Figure 6**. This migration is facilitated by significant relaxation and cooperative motion within the structure, particularly the rotational motion of the $GeO_4$ tetrahedra.

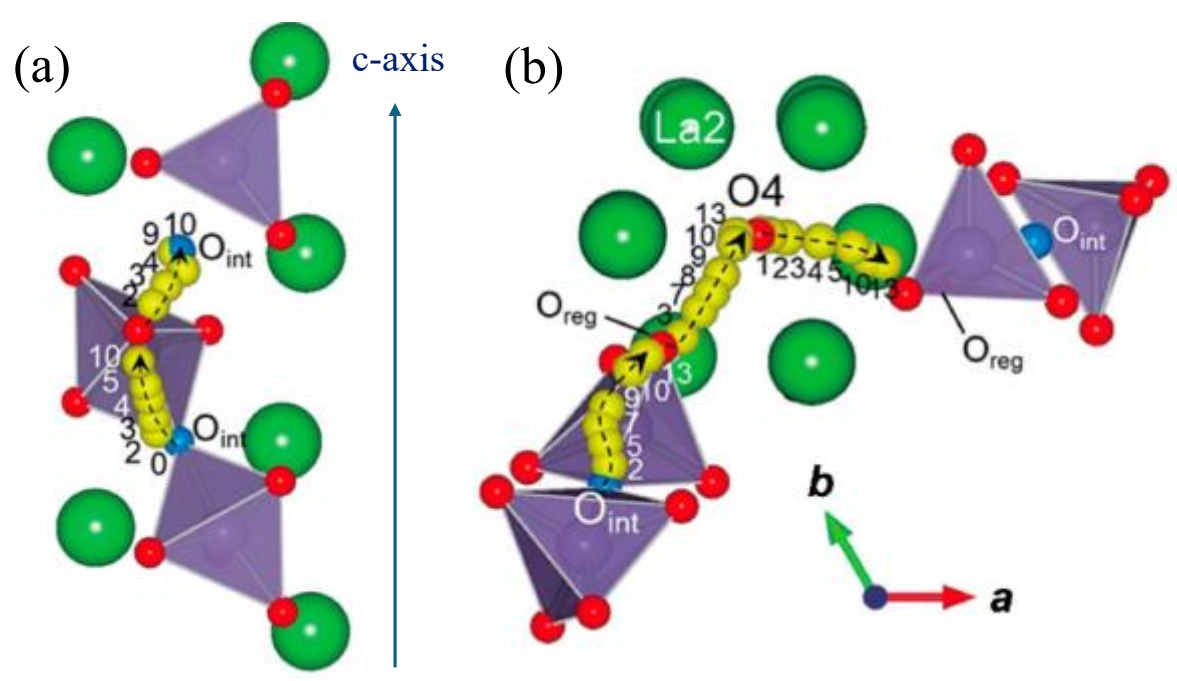

Figure 6 (a) Trajectories of interstitial oxygen ions (blue balls) diffuse along the c axis in $La_{10}Ge_6O_{27}$ by interstitialcy mechanism. (b) Trajectories of interstitial oxygen ions (blue balls) diffuse normal to the c axis by the “multi-oxygen ion cooperative mechanism”. Yellow balls demonstrated the diffusion pathways. Figures adapted with permission from Ref. [72].

Various substitutions have been explored in apatite, aiming to optimize their structure and enhance ionic conductivity. The apatite structure exhibits significant flexibility, accommodating diverse cation dopants at both La and Si/Ge sites. This structural adaptability enables substantial local distortions and variations in site volumes, making apatite structure particularly attractive for chemical substitution studies. Previous studies have shown that it is possible to introduce a wide range of dopants, e.g. $Zr^{4+}$, $Y^{3+}$, $Bi^{3+}$, $Sr^{2+}$, $Ba^{2+}$, $Mn^{2+}$, $Li^{+}$ onto the La site, and $Zn^{2+}$, $Mg^{2+}$, $Cu^{2+}$, $Ni^{2+}$, $Mn^{2+}$, $Co^{3+}$, $Al^{3+}$, $B^{3+}$, $Fe^{3+}$, $Ga^{4+}$, $Ti^{4+}$, $V^{5+}$, $As^{5+}$, $P^{5+}$ onto the Si/Ge site.[75,82,89–96] Numerous doping studies have been reviewed in the literature, showing that some dopants reduce conductivity, while others enhance it.[70,75,83,97–100] Here, we summarize the discussions and insights from previous reviews and highlight the most recent advancements. Key findings from earlier studies include: (1) In stoichiometric compositions such as $La_8M_2(SiO_4)_6O_2$ (M represents dopant on La sites), ionic conductivity is generally low due to the absence of interstitial oxygen. Doping strategies that generate oxygen-excess compositions tend to increase ionic conductivity by providing additional mobile charge carriers. (2) Dopants at the Si/Ge sites that facilitate pivoting, expansion, or distortion of the tetrahedra surrounding the conduction channels are critical for reducing migration barriers. Local structural studies have shown that this effect is related to the inherent flexibility of the silicate substructure, which permits relatively large local distortions and variations in site volumes.[100]

Continued doping investigations including co-doping strategies, stability enhancements, and detailed mechanistic studies have further optimized apatite-type materials for use as intermediate-temperature SOFC electrolytes. Among these efforts, Mg+Mo co-doping has been particularly effective, significantly improved both the ionic conductivity and densification properties of silicate apatite. The highest reported conductivity reached $3.39 \times 10^{-2}$ S/cm at 800 °C for $La_{9.5}Si_{5.45}Mg_{0.3}Mo_{0.25}O_{26+\delta}$, a significant improvement from $7.82\times10^{-3}$ S/cm for undoped lanthanum silicate oxides (LSO).[101] Phosphorus doping at the Si sites has also shown promise for enhancing ionic conductivity. The best conductivity achieved through P-doping was $7.27\times10^{-3}$ S/cm at 800 °C for $La_{9.33}Si_{6-x}P_xO_{26+x/2}$, attributed to an increased concentration of interstitial oxygen defects, which play a key role in enabling fast oxygen-ion transport.[95] Recent studies have explored doping with $Cr^{3+}$, $Fe^{3+}$, and $Mo^{6+}$ at the Si site in LSO, showing that these dopants can enhance ionic conductivity by modifying the oxygen channel structure and promoting interstitial oxygen migration. Electron paramagnetic resonance (EPR) studies indicated that Cr and Fe can introduce local lattice distortions, resulting in improved conductivity up to $3.33\times10^{-2}$ S/cm at 800 °C for Cr-doped LSO.[102]

Apatite oxides, particularly lanthanum silicate and germanate apatites, represent a class of interstitial oxygen conductors characterized by isolated tetrahedra and one-dimensional channels along the c-axis. These isolated tetrahedra and the channel not only host excess oxygen ions but also facilitate their migration via interstitial and interstitialcy mechanisms. The structure's inherent chemical tunability enables extensive aliovalent doping at both cation sites, which modifies local environments, stabilizes interstitials, and lowers diffusion barriers. Overall, the apatite structure offers a combination of isolated tetrahedra and diverse dopants adaptability, making it ideal for designing high-performance interstitial oxygen conductors.

### 2.5 Hexagonal perovskite derivative and Palmierite

Hexagonal perovskite derivatives $Ba_3MM'O_{8.5}$ (M = Mo, W; M'= Nb) have emerged as promising oxygen-ion conductors due to their unique layered structures and ability to accommodate interstitial oxygen since 2016.[103] $Ba_3MM'O_{8.5}$ crystallizes in a hybrid structure, combining 9R perovskite blocks with $(M/M')O_6$ octahedra and palmierite-type slabs composed of isolated $(M/M')O_4$ tetrahedra, as shown in **Figure 7a**. The presence of isolated $(M/M')O_4$ tetrahedra with non-bridging apical oxygen atoms is an important prerequisite for the ionic conduction observed in the $Ba_3MM'O_{8.5}$ system (M = W, Mo). A reduction in the ratio of tetrahedral to octahedral units diminishes the available fast diffusion pathways. For example, McCombie *et. al* observed that the strong reduction in the ratio of $(M/M')O_4$ tetrahedra to $(M/M')O_6$ octahedra results in a reduction in the ionic conductivity by an order of magnitude at 450 °C upon replacement of $W^{6+}$ for $Mo^{6+}$, indicating that the fast oxygen-ion diffusion pathways is most likely along the palmierite layers, attributed to the distortion of the flexible $(M/M')O_4$ tetrahedra.[104] Beyond Mo/W substitution, donor doping has proven highly effective. For instance, partial substitution of $Nb^{4+}$ by $V^{5+}$ in $Ba_3Nb_{0.9}V_{0.1}MoO_{8.5}$ leads to a high bulk ionic conductivity of ~ $10^{-2}$ S/cm at 600 °C, almost one order of magnitude higher than that of $Ba_3MNbO_{8.5}$.[105] Using *in situ* neutron diffraction and maximum-entropy method (MEM) analysis, Yashima *et al.*[106] revealed that the oxygen ions migrate two-dimensionally through mixed O2 and O3 sites in between the M1-O polyhedral layers in $Ba_3MoNbO_{8.5-\delta}$. Similar phenomenon was found in $Ba_3WNbO_{8.5}$, where the interstitialcy "push-pull" mechanism is invoked to explain two-dimensional oxide-ion transport.[107]

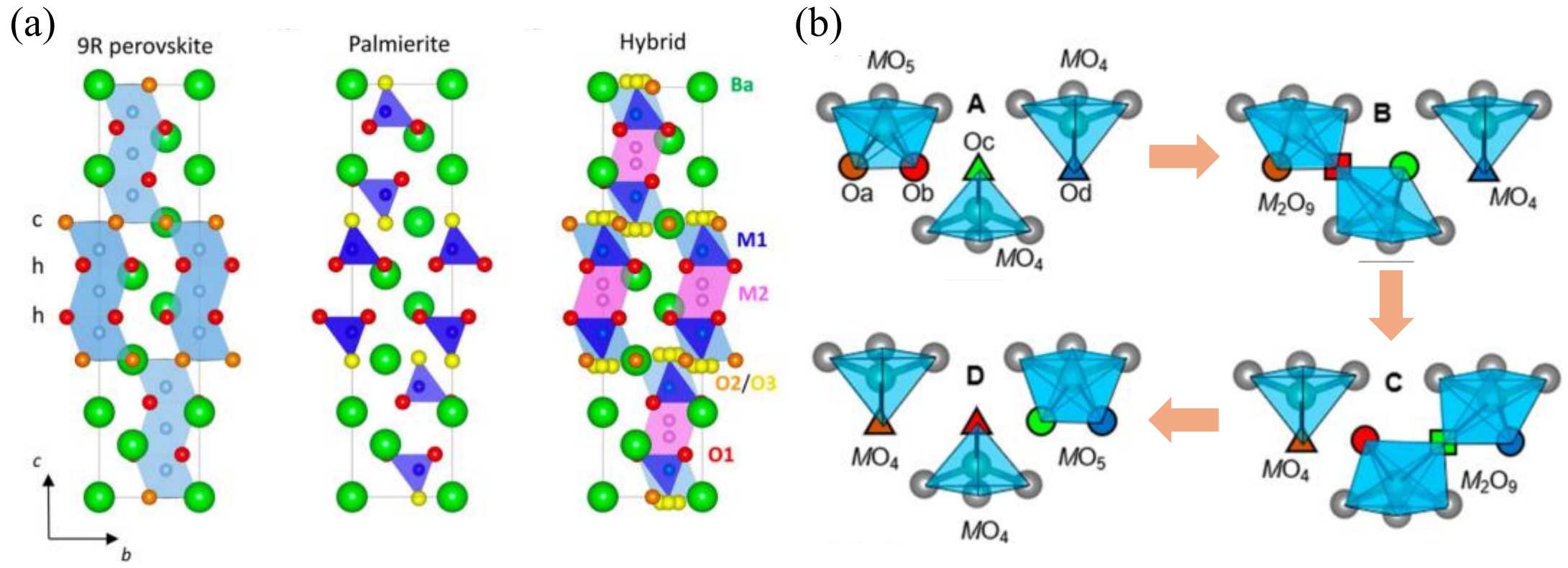

Figure 7 (a) 9R perovskite, palmierite, and hybrid crystal structure of $Ba_3NbMoO_{8.5}$. Blue and light blue polyhedra represent the average $M1O_x$ units created by partial occupation of O2 and O3. In the hybrid model, the M2 and O3 positions are split from the ideal sites to account for the cation and anion disorders. Figures adapted with permission from Ref. [105]. (b) Schematic representations of oxygen diffusion process of $Ba_7Nb_{3.8}Mo_{1.2}O_{20.1}$ at 1200°C. The brown (Oa), red (Ob), light green (Oc), and blue (Od) spheres represent mobile oxygen-ions in the oxygen-ion conducting c′ layer. Gray spheres denote the immobile oxygen ions forming the trigonal plane. The circle, triangle, and square shape represent an apical oxygen atom of $MO_5$, an apical oxygen atom of $MO_4$, and a corner-sharing oxygen atom in $M_2O_9$ dimer, respectively. Figures adapted with permission from Ref. [108].

Based on these findings, Fop *et al*. investigated a related hexagonal perovskite derivative $Ba_7Nb_4MoO_{20}$,[109] which is formed by an intergrowth of 12R perovskite blocks and palmierite layers. It exhibited a high oxygen-ion conductivity of $4\times10^{-3}$ S/cm at 510 °C, attributed to the relatively higher average concentration of lower coordination tetrahedra, indicating that the ionic conduction is facilitated by these isolated tetrahedra in the palmierite layers. Substitutional strategies employed various cation continue to enhance conductivity in these systems. Yashima *et al*. substituted $Nb^{4+}$ with $Mo^{6+}$ and obtained $Ba_7Nb_{3.9}Mo_{1.1}O_{20.05}$, which was found as a pure ionic conductor over a wide range of oxygen partial pressure range from $2\times10^{-26}$ to 1 atm at 600 °C. Its bulk conductivity reaches to $5.8\times10^{-4}$ S/cm at 310 °C, higher than $Bi_2O_3$ and zirconia-based materials.[38] Sakuda *et al*. substituted $Nb^{4+}$ with $Cr^{6+}$ and synthesized $Ba_7Nb_{4-x}Cr_xMoO_{20+x/2}$, which was found exhibit high oxygen ionic conductivity of $1.6\times10^{-3}$ S/cm at 508 °C and $1.1\times10^{-2}$ S/cm at 904 °C in static air, and wide electrolyte domain in the oxygen partial pressure $P(O_2)$ regions.[110] Murakami *et al*. prepared $Ba_7Ta_{4-x}Mo_{1+x}O_{20+x/2}$ (x = 0.2, 0.3, 0.5, 0.7) and found a large number of interstitial oxygen-ions in $Ba_7Ta_{3.7}Mo_{1.3}O_{20.15}$, leading to a high level of oxygen ionic conductivity of $1.08\times10^{-3}$ S/cm at 377°C, also with chemical and electrical stability in highly reduced atmospheres.[111] The high conductivity of these compounds was attributable to the interstitial-O5 oxygen diffusing two-dimensionally along the palmierite layers, similar to the migration paths in the $Ba_3MoNbO_{8.5-\delta}$. *Ab initio* molecular dynamics (AIMD) simulations on $Ba_7Nb_{3.8}Mo_{1.2}O_{20.1}$ indicated excess oxygen atoms are incorporated by the formation of both 5-fold coordinated $(Nb/Mo)O_5$ monomer and its $(Nb/Mo)_2O_9$ dimer with a corner-sharing oxygen atom.[108] The dynamic breaking and reformation of these dimers in oxygen-deficient layers enables rapid oxygen-ion migration, as illustrated in the diffusion pathway shown in **Figure 7b**.

Given that the interstitial formation and diffusion predominantly occur within the palmierite layers in hexagonal perovskite derivatives, it is natural to explore palmierite oxides directly as hosts for fast ionic transport. Tawse *et al*. synthesized the palmierite oxide $Ba_3Ti_{0.9}Mo_{1.1}O_{8.1}$ with the purpose of introducing interstitial oxygen into the palmierite structure via $Mo^{6+}$ substitution for $Ti^{4+}$.[40] Neutron diffraction confirmed that interstitial oxygen in the $[BaO_{2+x}]$ layer, as shown in **Figure 8.** The presence of interstitial oxygen resulted in mixed $Ti/MoO_x$ coordination (4-, 5-, and 6-fold coordinated polyhedral) and a more stable coordination environment for $Ti^{4+}$. The material exhibits a total conductivity of $3.9\times10^{-3}$ S/cm at 600 °C, enabled by the interstitialcy diffusion mechanism.

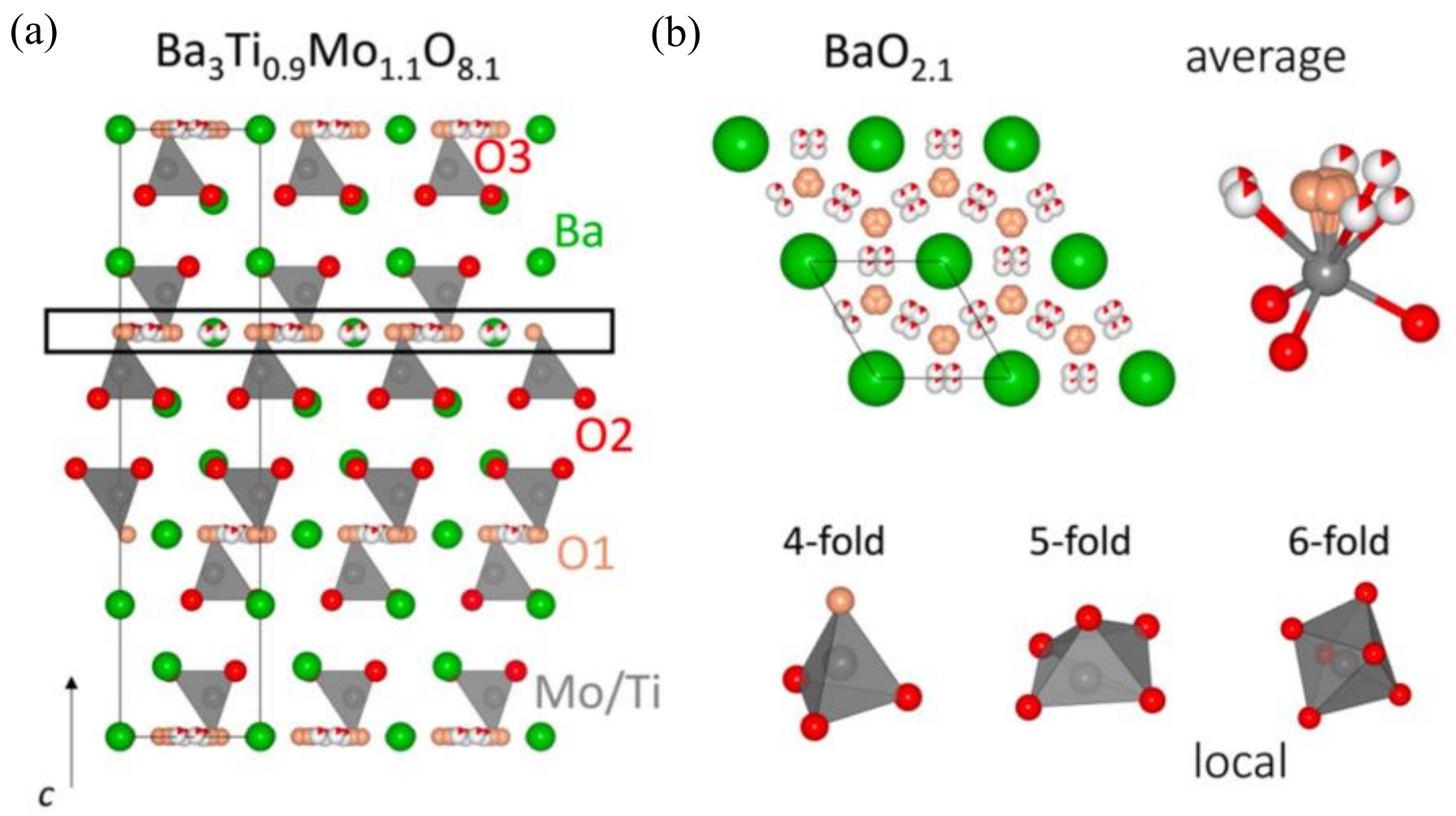

Figure 8 Average crystal structure of $Ba_3Ti_{0.9}Mo_{1.1}O_{8.1}$. (a) Crystal structure of $Ba_3Ti_{0.9}Mo_{1.1}O_{8.1}$, showing the isolated tetrahedral $MO_4$ units with interstitial oxygen defects. (b) [$BaO_{2.1}$] layer and representation of the average and local metal coordination in $Ba_3Ti_{0.9}Mo_{1.1}O_{8.1}$. The [$BaO_{2.1}$] layer is composed by partially occupied and split O1 and O3 positions, which leads to the formation of disordered metal polyhedral with average composition $MO_x$, resulting in 4-, 5-, and 6-fold coordination units on the local scale. Figures adapted with permission from Ref. [40].

These studies highlight that hexagonal perovskite derivatives and palmierite oxides form a versatile class of materials capable of accommodating and conducting interstitial oxygen-ions. A key structural feature in these materials is the presence of isolated tetrahedra arranged in palmierite layers, which introduce ample free volume and local structural flexibility for interstitial formation and diffusion. The diffusion of oxygen within these structures proceeds via a dimer-mediated cooperative mechanism, in which oxygen interstitials briefly pair with lattice oxygen to form dimers that facilitate fast ion conduction. This interstitialcy process significantly reduces the activation energy for diffusion, thereby enhancing the material's ionic conductivity. Furthermore, compositional engineering emerges as an effective design strategy for realizing high oxygen ionic conductivity in materials endowed with substantial structural flexibility.

### 2.6 Ruddlesden-Popper $La_2NiO_{4+\delta}$

Ruddlesden-Popper (RP) perovskite, which shared the tetragonal $K_2NiF_4$-type structure, features a layered arrangement consisting of perovskite-like blocks and rock-salt-like layers, as shown in **Figure 9a**. Due to its structural similarity to perovskite, which has been found as fast oxygen-ion conducting material via oxygen vacancies, Ruddlesden-Popper mostly is considered as potential vacancy-mediated oxygen-ion conductors for early studies. In the study of oxygen deficient $La_{2-x}Sr_xCuO_{4-y}$, it is revealed that the oxygen diffusion coefficient decreases as the concentration of oxygen vacancies increases, opposite to the typically trend in oxygen vacancy conductors but consistent with that in interstitial conductors.[112,113] Consequently, it is found that RP oxides can be both oxygen-deficient and oxygen-excess, and oxygen-ion diffusion can occur via mechanisms associated with either oxygen vacancies or oxygen interstitials.[114,115] Specifically, nickelates $La_2NiO_{4+\delta}$ has been found to exhibit a wide range of oxygen hyperstoichiometry ($0 \leq \delta \leq 0.25$),[116–119] with the incorporation of interstitial oxygen-ions into the rock salt layers, as shown in **Figure 9b**. Subsequent studies demonstrated that interstitial oxygen-ion diffusion is often more significant in Ruddlesden–Popper phases. For instance, Skinner and Kilner[120] employed isotope exchange depth profile method (IEDP) and found that oxygen-excess $La_2NiO_{4+\delta}$ exhibited higher oxygen diffusivity than $La_{0.6}Sr_{0.4}Co_{0.2}Fe_{0.8}O_3$ (LSCF) and only one order of magnitude lower than that of the best-performing perovskite $La_{0.3}Sr_{0.7}CoO_3$ (LSC). This high concentration of interstitial oxygen enables fast oxygen-ion conduction in $La_2NiO_{4+\delta}$, establishing Ruddlesden–Popper as a novel family of interstitial oxygen conducting materials.

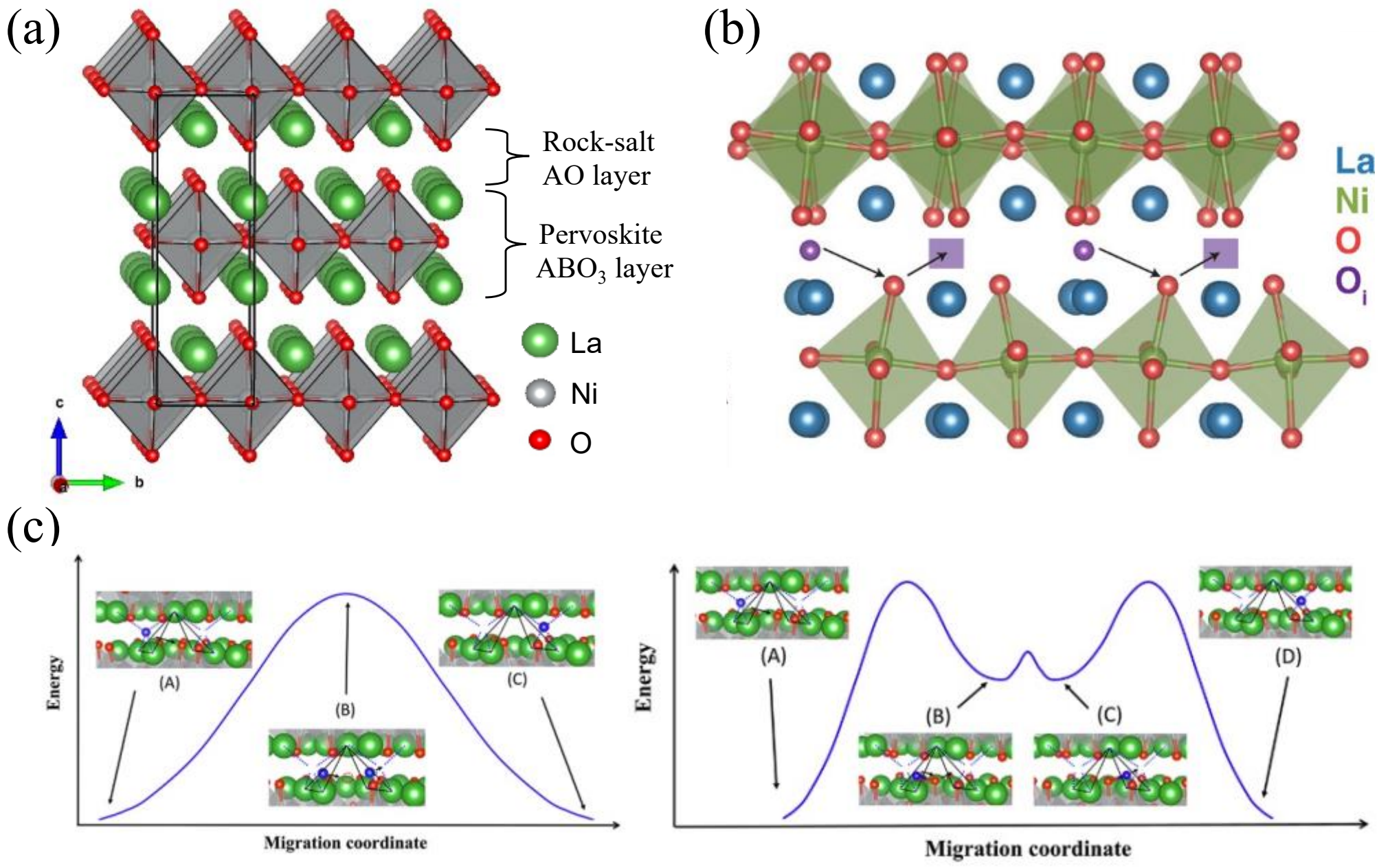

Figure 9 (a) Structure of Ruddlesden-Popper nickelates $La_2NiO_4$. (b) Snapshot displays an interstitial oxide ion migration via an interstitialcy mechanism in $La_2NiO_{4+\delta}$, $\delta = 0.125$. The migrating interstitial oxygen $O_i$ is depicted as a purple sphere, and the arrows show the direction of travel. The migrating oxide ion will displace the apical oxygen of the $NiO_6$ octahedron, which will move into the interstitial site denoted by the purple square. Note the tilting of the octahedra away from the interstitial. Figure adapted with permission from Ref. [121] (c) Diffusion pathway and the schematic energy landscape of the oxide−oxide (left) and oxide-peroxide (right) interstitialcy-mediated diffusion mechanism. The dashed blue lines show the tetrahedral coordination of $O_{int}^{2-}$. The blue curve is the schematic energy landscape. The green, silver, and red spheres denote La, Ni, and O atoms, respectively. The blue spheres denote the migrating O atoms. Figure adapted with permission from Ref. [122]

Further studies have revealed interstitial oxygen-ion diffusion pathways and anisotropic diffusion characteristics in RP oxides, as demonstrated by research employing neutron scattering experiment,[123] molecular dynamics (MD) calculation,[124] and density functional theory (DFT) studies.[121,125–127] The oxygen tracer diffusion coefficient was found highly anisotropic in $La_{2-x}Sr_xCuO_{4\pm\delta}$, with values in the a-b plane being two orders of magnitude higher than in the c direction.[114] Moreover, the diffusion coefficient decreases strongly with increasing Sr concentration, which was attributed to a shift in the diffusion mechanism from interstitial oxygen diffusion to oxygen vacancy diffusion with increasing Sr concentration. Similar anisotropic characteristics in oxygen exchange and transport properties were observed in $La_2NiO_{4+\delta}$ thin films,[128] $Pr_2NiO_{4+\delta}$ and $Nb_2NiO_{4+\delta}$ single crystals[129] experimentally using IEDP technique combining $^{16}O/^{18}O$ exchange and secondary ion mass spectroscopy (SIMS). MD simulation uncovered that the anisotropic diffusion in $La_2NiO_{4+\delta}$ and $Pr_2NiO_{4+\delta}$ is attributed to that the interstitial oxygen-ion diffusion occurs almost entirely via an interstitialcy mechanism in the a-b plane.[124,130] As shown in **Figure 9b**, this interstitialcy mechanism involves an interstitial oxygen displacing a neighboring apical oxygen, which in turn pushes the latter into another interstitial oxygen. This process has a significantly lower activation energy than migration via direct interstitial site exchange, according to DFT calculations. Furthermore, Xu et al.[122] identified two oxygen interstitialcy diffusion mechanisms involving oxide and peroxide interstitials in $La_{2-x}Sr_xNiO_{4+\delta}$ using DFT studies. As illustrated in **Figure 9c,** the oxide–oxide mechanism (left) involves an oxide interstitial ($O_{int}^{2-}$) displacing the apical oxygen atop "kicks out" the apical oxygen at the top of the $NiO_6$ octahedron, creating a temporary vacancy. In contrast, the oxide-peroxide mechanism (right) involves a peroxide interstitial ($O_{int}^{-}$) bonds with the apical oxygen to form an $O_2^{2-}$ dumbbell, without generating oxygen vacancies. Under relatively reduced conditions (Ni valence ~2.25), the oxide–oxide mechanism dominates, exhibiting a low activation energy of ~0.7 eV. As the system becomes more oxidized (Ni valence ~2.5), the oxide–peroxide mechanism becomes competitive, and both mechanisms exhibit higher activation energies of ~1.6–1.8 eV.

Many efforts have been made to understand the factors governing the formation and migration energy of interstitial oxygen in Ruddlesden–Popper perovskites. Studies of lattice dynamics have highlighted the crucial role of epitaxial strain, dynamical delocalization of apical oxygen atoms, and facile octahedral rotations, all of which contribute to lowering the migration barriers.[121,122,126] DFT studies found that a modest epitaxial tensile strain of 2% results in a slight reduction in migration barriers in $La_{2-x}Sr_xNiO_{4+\delta}$ across most typical Ni valence values.[122] In $Nd_2NiO_{4+\delta}$, excess oxygen on interstitial lattice sites activates large displacements of the apical oxygen atoms of the corner-sharing $NiO_6$ octahedra along the [110] direction, thus favoring their diffusion within the rock-salt layer via an interstitialcy mechanism. This highlighted that dynamical delocalization of apical oxygen atoms along [110] is necessary to facilitate ionic mobility at room temperature.[126] First-principles calculations even quantified the microscopic link between octahedral rotation modes in $Ln_2NiO_{4+\delta}$ (Ln=La, Pr, Nd), the results shown that migration barriers can be correlated with the tendency of the material to undergo octahedral rotation distortion, where crystal chemistry and epitaxial strain can be harnessed as design strategies to lower these barriers.[121] Key structural features of flexible apical oxygen positions associated with corner-sharing octahedra, dynamic octahedral tilting, and strain-sensitive lattice dynamics contribute to low migration barriers and high oxygen diffusivity. Despite the fact that ionic transport is a complex physical process and can be affected by many factors, these fundamental understanding of oxygen transport in Ruddlesden−Popper perovskites has advanced effective material design strategies for engineering RP oxides with higher ionic conductivity.

### 2.7 Melilite

Melilite-type materials have emerged as promising candidates for solid oxide fuel cells (SOFCs) applications due to their notable oxygen-ion conductivity at intermediate temperatures (500-900°C), primarily facilitated by interstitial oxygen diffusion. In 2004, Rozeumek *et al.*[131] synthesized $La_{1+x}Sr_{1-x}Ga_3O_{7-\delta}$ and performed four-point direct-current conductivity measurements, revealed that this material exhibited p-type behavior in air, with conductivity reaching approximately 0.1 S/cm at 950°C. This finding was pivotal in demonstrating the role of oxygen-ion in enhancing conductivity. Kuang *et al.* subsequently conducted an in-depth investigation of the La-rich composition $La_{1.54}Sr_{0.46}Ga_3O_{7.27}$, which exhibited a higher conductivity of 0.02–0.1 S/cm over the temperature range of 600-900 °C.[42] The melilite structure consists of alternating layers of corner-sharing $GaO_4$ tetrahedral network and a layer of $La^{3+}/Sr^{2+}$ cations located above the five-ring centers. **Figure 10a-b** display the structure and accommodation of oxygen interstitials in melilite $La_{1.54}Sr_{0.46}Ga_3O_{7.27}$. Within the tetrahedral network, two distinct types of tetrahedra exist. The Ga1-centered tetrahedron (blue) connects to four neighboring Ga2-centered tetrahedra (green), while the Ga2-centered tetrahedra connect to two Ga1-centered tetrahedra and one Ga2-centered tetrahedron, all by sharing corners. This corner-sharing tetrahedral network provided the structural capacity to host interstitial oxygen ions in the space within these pentagonal $GaO_4$ rings in the a-b plane, as well as in between La/Sr cations along the c-axis, as represented by the O4 site in **Figure 10ab**. This study marked a key advancement in understanding that the interstitial oxygen serves as the primary charge carriers. Through Rietveld refinement of neutron powder diffraction (NPD) data and detailed analysis, the mechanisms of stabilization and migration of interstitial oxide defects were elucidated.

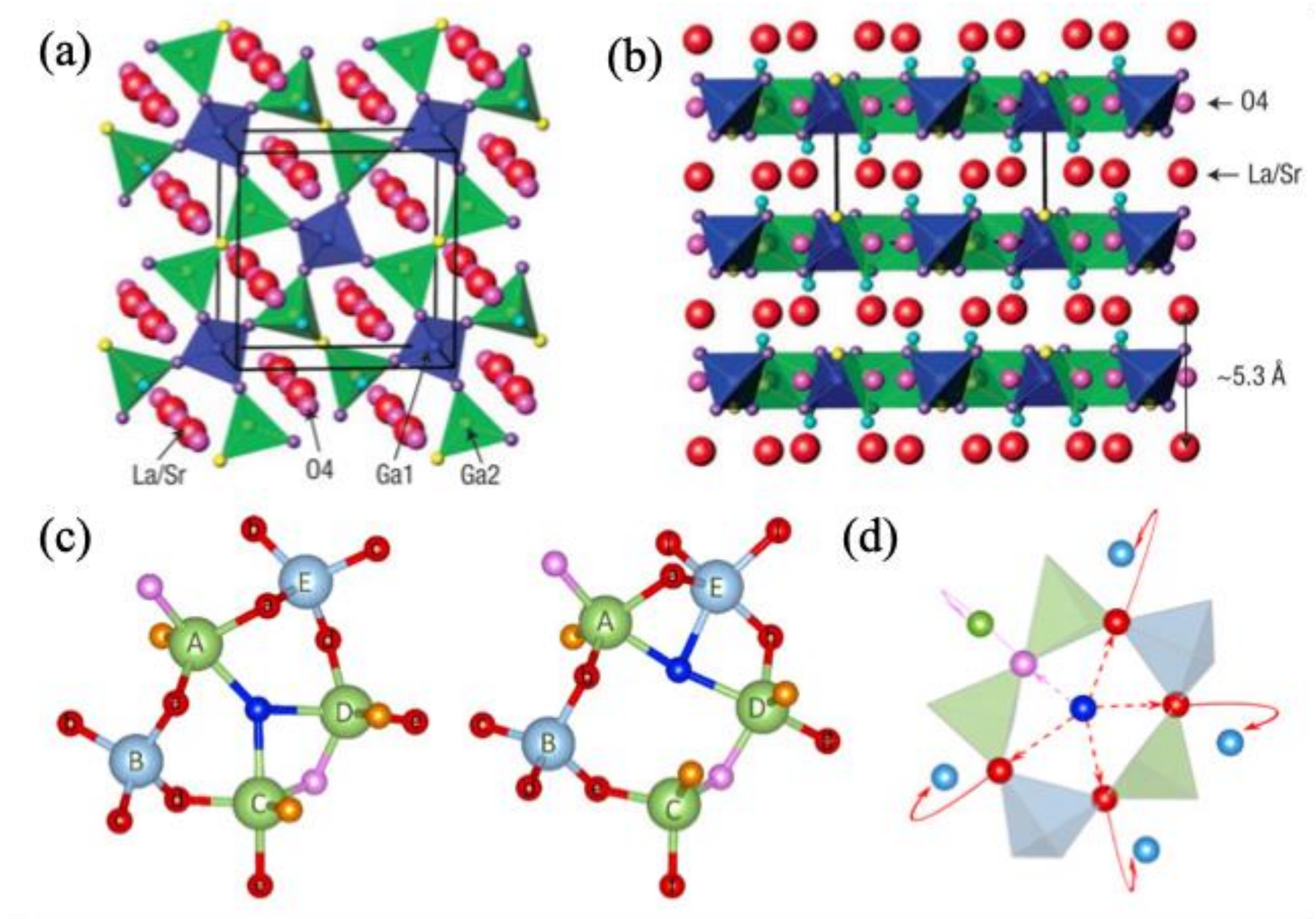

Figure 10 (a, b) Structure and accommodation of oxygen interstitials in melilite $La_{1.54}Sr_{0.46}Ga_3O_{7.27}$. (a) 2D $Ga_3O_7$ corner-shared tetrahedral layers with La/Sr columns (red) and interstitial oxygen O4 (pink) located in the distorted pentagonal tunnels viewed along the c-axis direction. (b) The stacking of the $(La/Sr)_2$ and $Ga_3O_7$ structural units along the c axis shows that the O4 interstitials are located in the $Ga_3O_7$ layers. The two distinct Ga-centred tetrahedra are distinguished by coloring. (c) Relaxed structures of $La_{1.5}Sr_{0.5}Ga_3O_{7.25}$. Oxygen interstitial $O_i4$ is located in the center (left) and oxygen interstitial $O_i5$ is located off-center (right) of the pentagonal ring formed by $GaO_4$ tetrahedra. Key: Ga1 (light blue), Ga2 (green), O1 (pink), O2 (orange), O3 (red), and $O_i$ (dark blue). (d) Five possible migration pathways of the oxygen interstitial within the a-b plane. Only two distinct migration paths exist, one across the O3 ion (red path) and one across the O1 ion (pink path). The $O_i$ moves to the lattice position (dotted arrows). The lattice oxygen ion moves to the next interstitial position (solid arrows). Key: initial $O_i$ (dark blue), final $O_i$ (blue and green), $(Ga1)O_4$ tetrahedron (blue), and $(Ga2)O_4$ tetrahedron (green). Figures adapted with permission from Ref. [42,132].

*Ab initio* calculations further revealed that interstitials in La-rich compositions preferentially occupy two stable positions within the tetrahedral network layer, either at the center of the pentagonal $GaO_4$ rings or at slightly off-centered position depending on the local cation environment, as shown in **Figure 10c**. The ion migration follows an interstitialcy mechanism hopping between adjacent pentagonal $GaO_4$ rings with relatively low barriers of 0.15 and 0.35 eV, as shown in **Figure 10d**.[132] This diffusion is facilitated by the flexibility of the tetrahedral network, which allows for rotation and deformation of the tetrahedra. Due to the layered structural feature, it is straightforward that oxygen diffusion in melilite is anisotropic. Wei *et al.*[133] studied melilite $[A_{1+x}B_{1-x}]_2[Ga]_2[Ga_2O_{7+x/2}]_2$ ($0 < x < 0.5$) (A=La, Nd; B=Ca, Sr) and found that intralayer oxygen-ion conduction is dominant in large single crystals. Conductivity measurements at 850°C revealed that conductivities were approximately $8\times10^{-3}$ S/cm along the c-axis and $3.6\times10^{-2}$ S/cm in a-b plane.

To further enhance performance, doping strategies have been explored. Studies on different dopants substitutions on A sites shown that the size of cations plays a pivotal role in stabilizing interstitial oxygen defects. Liu *et al.* demonstrated that lanthanide cations smaller than $La^{3+}$ are much less favorable for stabilizing the interstitial oxygen-ions in $Ln_{1+x}Sr_{1-x}Ga_3O_{7+0.5x}$ from the experimental synthesis point of view.[134,135] While for $La_{1+x}M_{1-x}Ga_3O_{7+0.5x}$, smaller cation dopant $Ca^{2+}$ tends to stabilize more interstitial oxygen-ions than $Sr^{2+}$ and $Ba^{2+}$.[136] These results suggested that there is an optimal cation size combo in melilite that best balance the expansion and contraction of pentagonal rings to allow efficient interstitial oxygen-ion stabilization and migration, as realized in $La_{1.54}Sr_{0.46}Ga_3O_{7.27}$, achieved the highest ionic conductivity.[42] A recent advancement involves substituting $Sr^{2+}$ with $Bi^{3+}$ in $LaSr_{0.6}Bi_{0.4}Ga_3O_{7.2}$ has further increased the ionic conductivity. The $Bi^{3+}$ ions successfully occupy the La/Sr sites in the lattice, causing lattice contraction and introducing interstitial oxide ions. DFT calculations have shown that the formation energy of these interstitial oxygen-ions is low, leading to significant conductivity improvements, reaching to ~$1.62\times10^{-2}$ S/cm at 800°C.[137]

In summary, the unique structural features of melilite-type materials, particularly their capacity to host interstitial oxygen ions within the corner-sharing tetrahedra network, are critical to their function as oxygen-ion conductors. The network's polyhedral frameworks, along with large electropositive cations that readily change coordination number, provide the structural flexibility necessary to accommodate interstitial oxygen ions and facilitate their conduction.[42]

### 2.8 Hexagonal manganite h-$RMnO_3$

Hexagonal manganite h-$RMnO_3$ (R = Sc, Y, Ho–Lu), initially attracted attention for their multiferroic properties and magnetoelectric coupling. However, their defect chemistry remained largely overlooked until 2010s, when studies began to uncover the pivotal role of interstitial oxygen in modulating their electrical conductivity. As shown in **Figure 11a**, the h-$RMnO_3$ (space group *$P6_3cm$*) features layers of corner-sharing $MnO_5$ trigonal bipyramids. These $MnO_5$ units are tilted in a trimerized pattern, while planes of $R^{3+}$ (e.g. $Y^{3+}$) ions lie between the $MnO_5$ layers, displaced along the c-axis due to the tilt. The hexagonal manganite structure is ~11% less dense than its orthorhombic perovskite counterpart, resulting in increased free volume suitable for hosting interstitial oxygen ions.

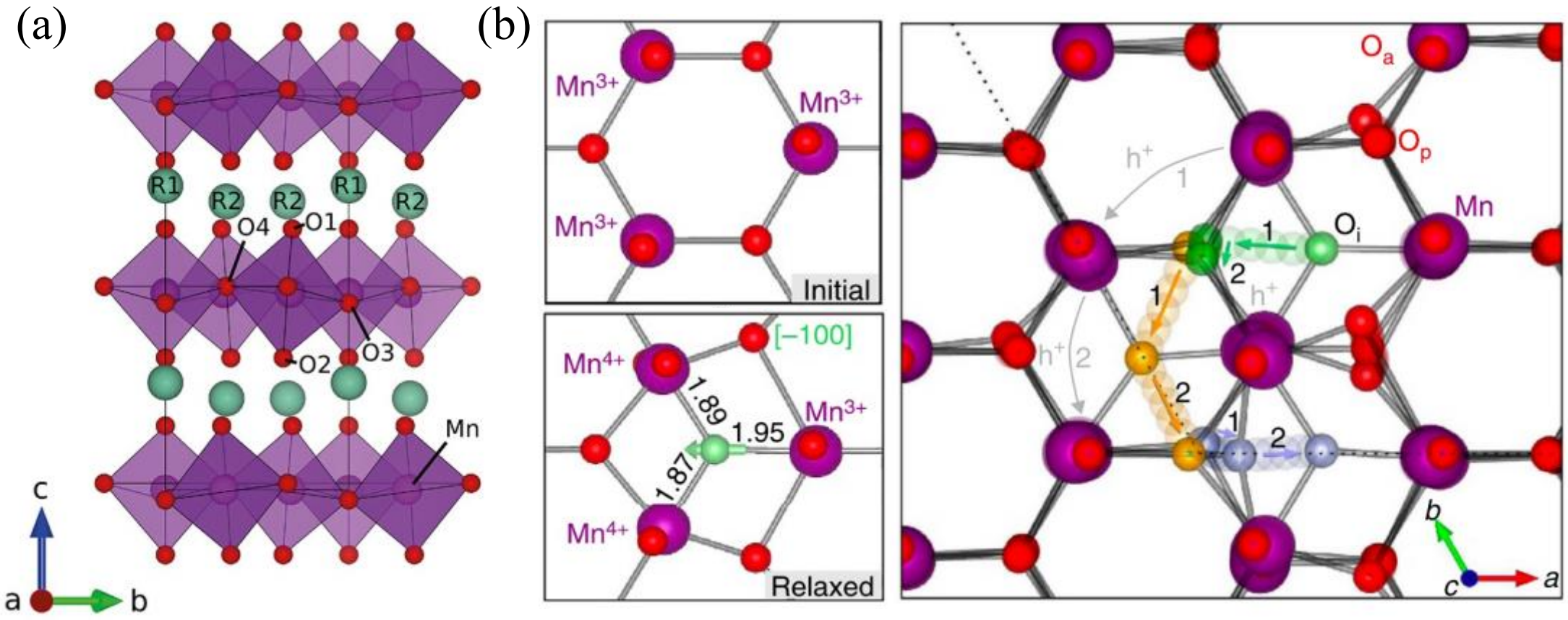

Figure 11 (a) Hexagonal P63cm crystal structure with turquoise $R^{3+}$ cations, red oxygen atoms, and the $MnO_5$ trigonal bipyramids shown as purple polyhedral. (b) Mn-O plane seen along the c axis, with the added $O_i$ (green) on the interstitial site in between Mn atoms. Left upper: unrelaxed structure; left lower: relaxed structure. The green arrow on $O_i$ indicates the displacement towards the two $Mn^{4+}$ ions. Bond lengths between $O_i$ and Mn are given in Å. Right: Subsequent migration paths of $O_i$ (green, then yellow and finally blue) through an interstitialcy mechanism nudging first a planar lattice oxygen at an O4 site (path 1) and then a planar lattice oxygen at an O3 site (path 2). The hopping of holes ($h^+$) between Mn sites are shown by grey arrows. Figures adapted with permission from Ref. [138,139]

A key discovery came from studies of $Dy_{1-x}Y_xMnO_{3+\delta}$, which revealed a large reversible excess oxygen content with δ up to 0.35 under air and oxygen atmospheres at low temperatures of 200 to 400°C, highlighting the material's exceptional oxygen storage capacity.[140] However, the energetics and kinetics of oxygen defects in h-$RMnO_3$ have not been addressed until 2016. Skjærvø *et al.*[43] thoroughly mapped out the energetics and the local structural distortions of excess oxygen incorporation in $YMnO_3$. Systematic TGA, electrical-conductivity, thermopower data, combined with first-principles calculations have revealed that oxygen is inserted into interstitial sites in the Mn-O planes. In **Figure 11b**, $O_i$ forms two shorter and one longer Mn-O bonds after relaxation, displacing toward two oxidized $Mn^{4+}$ ions and away from a $Mn^{3+}$ ion. This oxidation leads to the creation of mobile holes, giving rise to p-type conductivity. $O_i$ diffuses via an interstitiacy mechanism where it pushes an adjacent planar oxygen into a neighboring interstitial site, with moderate energy barriers of 0.48-0.62 eV from NEB calculations.

Substituting $Mn^{3+}$ with $Ti^{4+}$ tends to further stabilize and create extra interstitial oxygen. Danmo *et al.*[139] studied on $YMnO_3$ and $DyMnO_3$ and shown that $Ti^{4+}$ doping raises the maximum oxygen content and promotes absorption at lower temperatures. $Ti^{4+}$ doping expands the ab-plane slightly and lowers the energy barrier for $O_i$ migration, indicating more rapid and reversible oxygen exchange. In a recent study from Danmo and co-workers,[138] they introduced multiple rare-earth cations in single-phase hexagonal manganite and found that these high-entropy hexagonal manganite could stabilize the layered framework and maintain large, rapidly accessible oxygen storage even at moderate temperatures, exhibiting high oxygen surface kinetic. This study implies that structural disorder itself contributes to the improved oxidation kinetics. Improved properties and insensitivity to exact rare earth composition enhance the application potential of hexagonal manganite for oxygen separation and release.

### 2.9 Mayenite $Ca_{12}Al_{14}O_{33}$

Mayenite ($Ca_{12}Al_{14}O_{33}$ or $12CaO \cdot 7Al_2O_3$, C12A7) as a stable ceramic material, is first noted for its occurrence in cement clinkers.[141] It has attracted considerable attention since its initial identification as a high oxide-ion conductor in the late 1980s. Early research by Lacerda and co-workers demonstrated that the bulk ionic conductivity of mayenite is around $10^{-3}$ S/cm at ~1000 °C, which is only one order of magnitude lower than that of YSZ.[44] These findings prompted widespread interest in understanding mayenite's crystal structure, defect chemistry, and conduction mechanisms. Mayenite crystallizes in a cubic structure (space group $I\bar{4}3d$), comprising aluminate frameworks and $Ca^{2+}$ ions, with 12 sub-nanometer cages per unit cell. As shown in **Figure 12a**, $Ca_{12}Al_{14}O_{33}$ contains a three-dimensional network of eight-membered rings of $AlO_4$ tetrahedra linked via corner-sharing. 32 oxygens in the formula unit belong to the aluminate framework, while the additional oxygen ion randomly distributed within the cages (**Figure 12b**). These cage-entrapped oxygen ions are loosely bound and effectively act as interstitial species, capable of migrating between cage sites. As a result, mayenite exhibits relatively high oxide-ion mobility, with the cage-based oxygen ions providing the structural basis for the material's notable ionic conductivity.

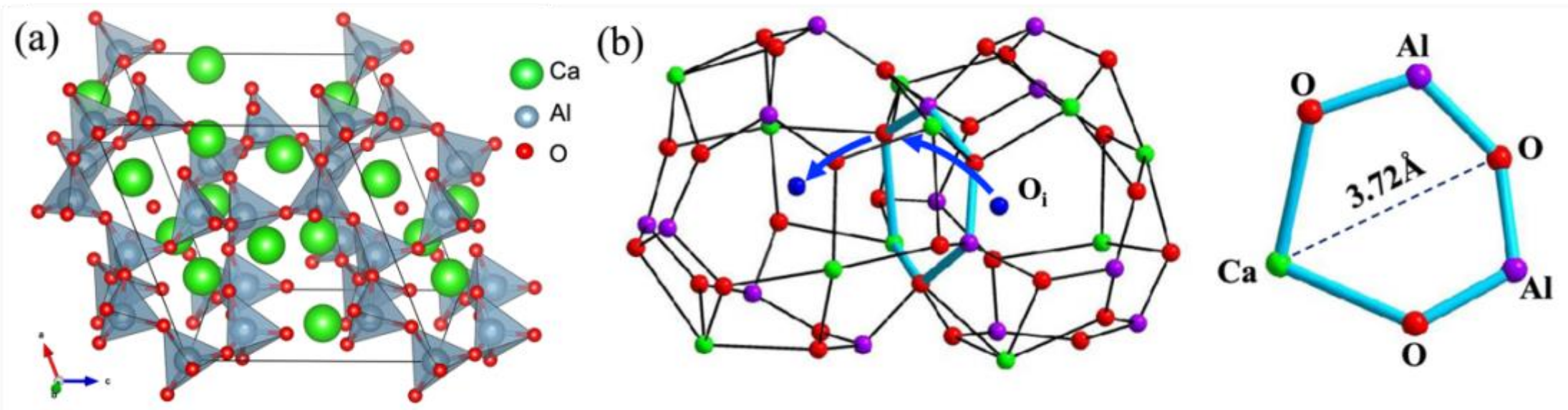

Figure 12 (a) Atomic structure of Mayenite $Ca_{12}Al_{14}O_{33}$. (b) Two neighboring cages sharing the ~3.72 Å wide window in the mayenite crystal structure. The green, purple, red, and blue spheres represent Ca, Al, framework O, and free O atoms, respectively. The blue dashed arrows indicate the interstitialcy diffusion pathways, serving as a visual guide to highlight how the cage-entrapped interstitial oxygen-ions exchange with framework oxygens. Figures adapted with permission from Ref. [142].

Shortly after the oxygen ionic conductivity reported $Ca_{12}Al_{14}O_{33}$ by Lacerda *et al.*,[44] researchers sought to clarify whether interstitial oxygen-ions in mayenite diffuse purely as classical interstitials or via an alternative exchange mechanism. Two distinct pathways were proposed: "free" $O^{2-}$ ions diffuse directly through cage windows via an interstitial mechanism or displace framework oxygens via an interstitialcy mechanism. Investigations using raman spectra analysis, oxygen isotope exchange, and *ab initio* calculations revealed that $O^{2-}$ diffusion occurs via interstitialcy mechanism between $O^{2-}$ in the cage and oxygen ions constituting the cage framework shown in **Figure 12b**, as the interstitialcy mechanism exhibits a comparatively lower migration barrier of approximately 0.58 eV, whereas the straightforward interstitial diffusion mechanism features a higher barrier of around 1.2 eV.[143,144]

Despite mayenite's favorable intrinsic properties for oxygen-ion conduction, its overall conductivity at moderate temperatures remains below that of the leading oxide electrolytes. In efforts to address this limitation, researchers have explored doping strategies aimed at tuning the cage geometry, carrier concentration, and overall defect equilibria. Early attempts involved substituting $Fe^{3+}/Co^{3+}$ for $Al^{3+}$ sites, in the hope of enlarging cage windows or modifying bonding interactions.[145] However, such interventions tended to reduce the bulk conductivity rather than enhance it. Recent studies indicate that $Ga^{3+}$ substitution can be more effective, given its larger ionic radius that may expand the sub-nanometer cage windows and facilitate oxygen transport Specifically, Yi *et al.* demonstrated that moderate Ga doping (e.g., up to x ≈ 0.4 in $Ca_{12}Al_{14-x}Ga_xO_{33}$) yields slightly higher bulk conductivities relative to pure mayenite.[146]

It is worth noting that a unique feature of mayenite is its capacity to stabilize a broad spectrum of species in its cage framework, including electrons. Under reducing atmospheres, it transforms into an inorganic electride (C12A7:$e^-$), with electrons occupying cage sites as anionic electron centers. Under oxidizing conditions, radicals such as $O^-$ or $O_2^-$ may reside in the cages,[147] either enhancing or complicating conduction pathways. This versatility in accommodating different redox states highlights the exceptional structural flexibility of the sub-nanometer cages and has led to extensive research efforts into tailoring mayenite's electronic and ionic conductivities for applications spanning catalysis and electrochemical devices. Ongoing studies employing advanced characterization techniques and computational modeling are expected to further harness the interstitial nature of mayenite to optimize its ionic conductivities and expand its functionalities in catalysis and redox-based devices.

### 2.10 Langasite

Langasite $La_3Ga_{5-x}Ge_{1+x}O_{14+x/2}$ has recently been recognized as an interstitial oxygen-ion conductor due to its three-dimensional framework of corner-shared tetrahedra and octahedra, which can accommodate extra oxygen in the interstitial sites,[45] although it was known as a oxygen vacancy conductor early in 2006.[148] As shown in **Figure 13ab**, langasite features 3D framework of mixed $GaO_6$/$GaO_4$ coordination, offering flexibility for structural modifications. Neutron diffraction studies on substituted langasites $La_3Ga_{5-x}Ge_{1+x}O_{14+x/2}$ have shown that interstitial oxygen ions are stabilized in corner-sharing polyhedral pairs, yielding (Ga,Ge)$O_8$ units, as shown in **Figure 13d**. Although langasite can accommodate a higher density of interstitials compared to melilite, the newly formed (Ga,Ge)$O_8$ edge-sharing motifs tends to trap the interstitial oxygen-ions, thereby suppressing long-range diffusion. Despite this limitation, the ability to incorporate substantial oxygen excess without compromising structural stability makes langasite-type materials attractive for the development of intermediate-temperature solid oxide fuel cell (IT-SOFC) electrolytes.

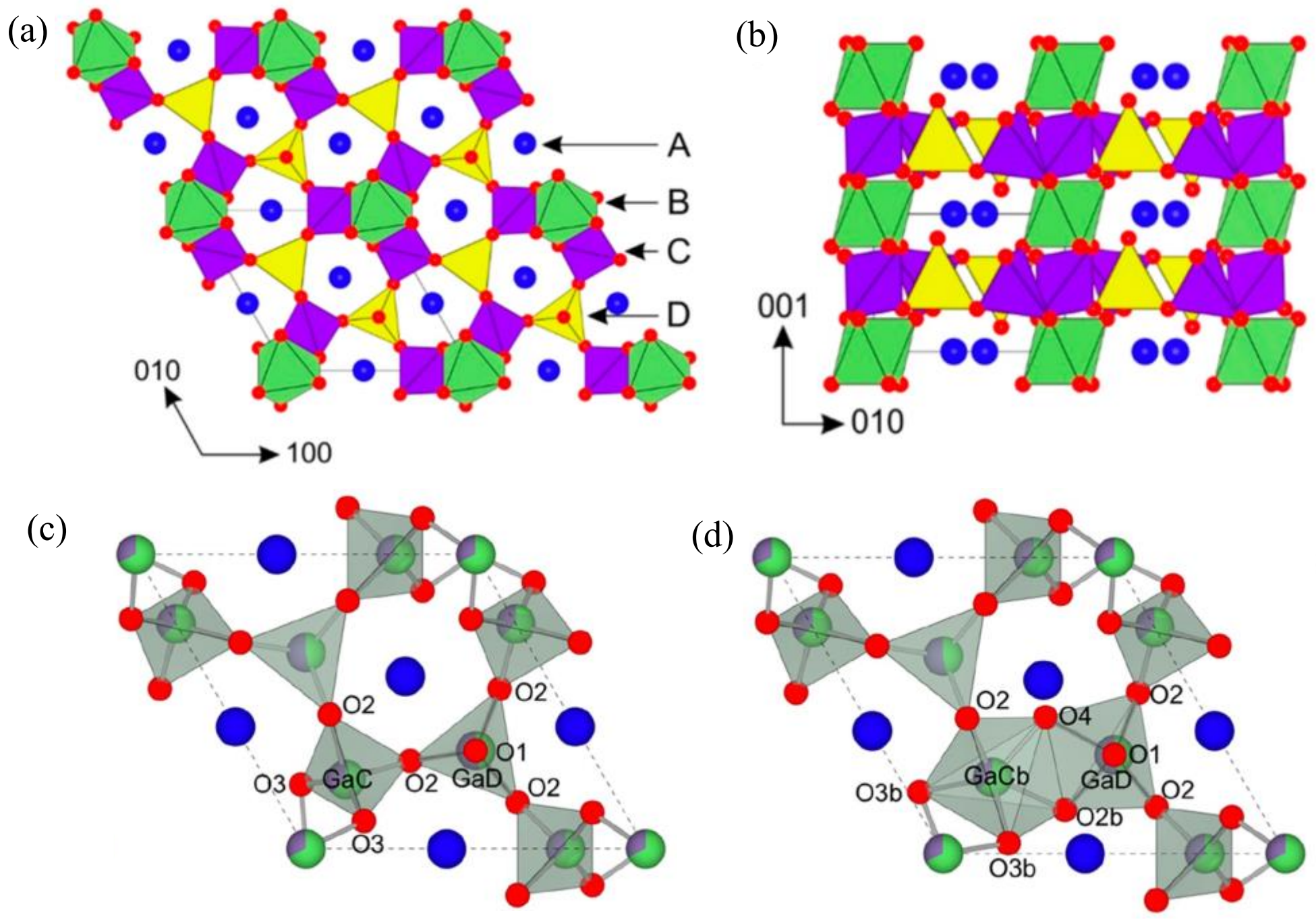

Figure 13 The langasite structure $A_3BC_3D_2O_{14}$ viewed along the stacking axis (a) and perpendicular to the stacking axis (b), showing a two-dimensional arrangement of three-connected and four-connected vertex-linked tetrahedra, which are connected along the stacking axis by vertex-linked $BO_6$ octahedra (green) to produce a three- dimensional network with six-membered ring channels. (c) The configuration of a noninterstitial-containing unit cell $La_3Ga_{3.5}Ge_{2.5}O_{14.75}$ from Rietveld model. (d) One possible local configuration of an interstitial-containing unit cell of $La_3Ga_{3.5}Ge_{2.5}O_{14.75}$, constructed from the Rietveld model, where O4 is the interstitial oxygen located in between GaC-O4 and GaD-O4 tetrahedra. Figures adapted with permission from Ref. [45].

Importantly, langasites have also demonstrated oxygen vacancy-mediated conduction when acceptor-type dopants (e.g., $Sr^{2+}$ substitution for $La^{3+}$, $Ga^{3+}$ substitution for $Ge^{4+}$) were introduced to generate additional vacancies in the oxide sublattice.[148] Similar to their interstitial counterparts, these vacancies are stabilized by the flexible coordination environment of Ga/Ge cations, and their mobility is facilitated by the interconnected tetrahedral and octahedral framework. Oxygen vacancies can coexist with electronic charge carriers, leading to mixed ionic-electronic conduction, the ability to control the content and distribution of vacancies in langasite structures provides another route for optimizing oxygen-ion conductivity.[148,149] These studies demonstrated that langasites offer a promising avenue for the development of high-performance oxide ion conductors due to their unique structural features and defect chemistry.

### 2.11 Garnet

While garnet oxides have traditionally been studied for their lithium-ion conductivity, recent developments have drawn attention to their oxygen-ion conduction capabilities.[46,47,150,151] Compounds such as $Ca_3Fe_2Ge_3O_{12}$ have demonstrated promising oxygen-ion transport, with conductivity values approaching those of benchmark materials like YSZ. Notably, the oxygen conductivity of donor-doped $Ca_{2.7}La_{0.3}Fe_2Ge_3O_{12}$ has been reported to reach approximately $10^{-3}$ S/cm at 700 °C.[150] The crystal structure of $Ca_3Fe_2Ge_3O_{12}$, as shown in **Figure 14a**, adopts a cubic symmetry with the space group *Ia-3d*. Within this structure, $Fe^{3+}$ occupies the octahedral (6-fold) coordination, $Ge^{4+}$ occupies tetrahedral (4-fold) coordination, and $Ca^{2+}$ is coordinated by an average of ~7.8 oxygen atoms. All oxygen atoms are located on the 96h Wyckoff sites, contributing to the three-dimensional connectivity of the oxygen sublattice. In the garnet structure, the polyhedra are connected primary through corner-sharing, which is a key structural feature for structural flexibility and ion transport.

Lee et al. investigated the formation energy and migration barrier energy for oxygen vacancies and interstitials in $Ca_3Fe_2Ge_3O_{12}$ by first-principles calculations.[46] They found that interstitial oxygen diffusion via the “kick-out” mechanism exhibited significantly lower migration barriers (0.19-0.45 eV) compared to vacancy-mediated diffusion (0.76 eV). Two main diffusion pathways of interstitial oxygen-ion were identified in **Figure 14b**: (1) direct interstitial hopping along 48g-96h′ sites, and (2) an interstitialcy mechanism involving 96h′-O-96h′ hopping. The latter enables 3D and delocalized long-range diffusion of interstitial oxygen-ion, leading to enhanced ionic conductivity in garnet oxides.

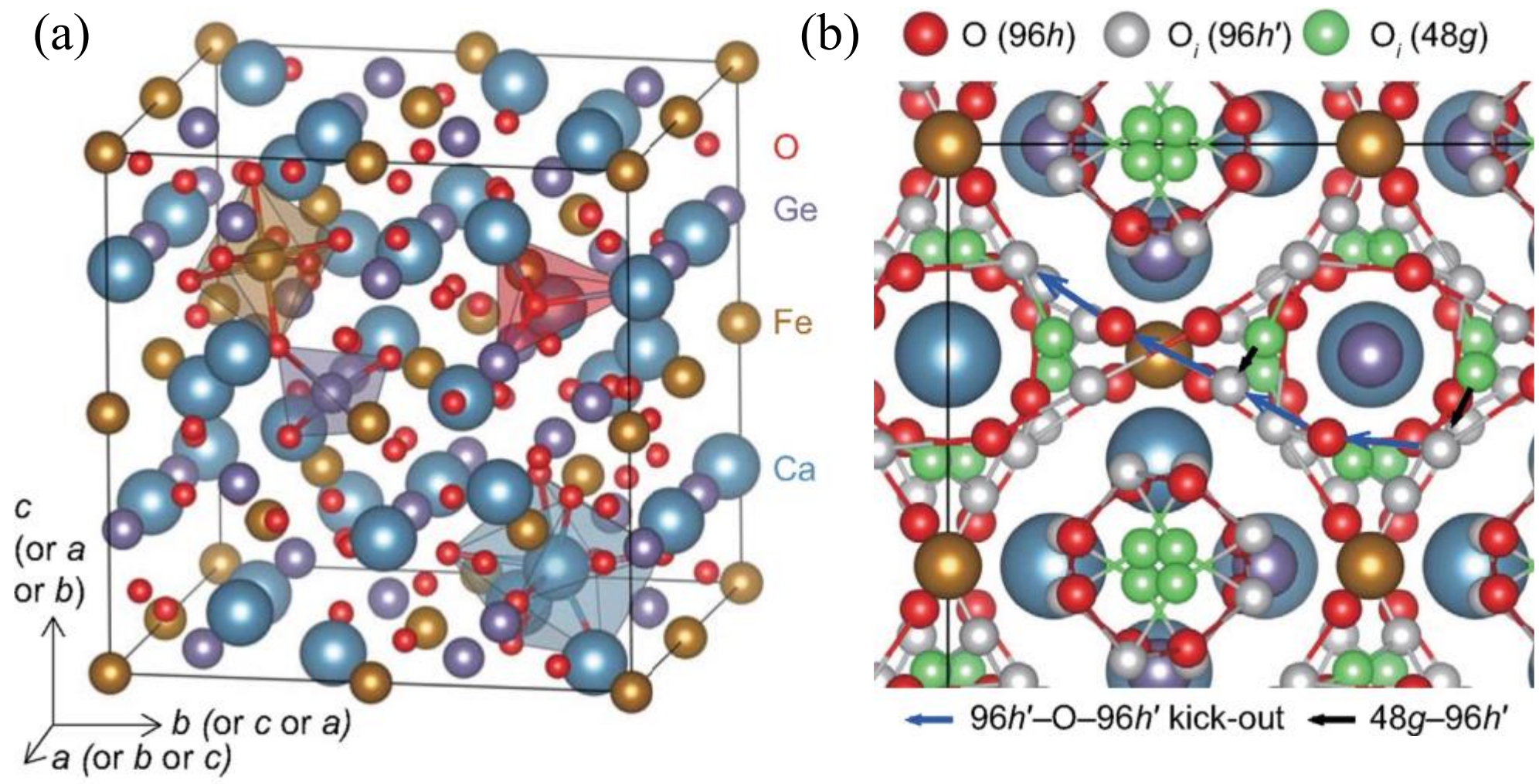

Figure 14 (a) Garnet crystal structure (space group: Ia-3d) of $Ca_3Fe_2Ge_3O_{12}$; the conventional cell with the cubic structure ($Ca_{24}Fe_{16}Ge_{24}O_{96}$, 160 atoms) is shown. (b) Distribution of equivalent interstitial oxygen $O_i$ sites 48g and 96h′ sites after the optimization. Two main migration pathways of $O_i$ are present: interstitialcy mechanism for 96h′–O–96h′ migration path, and interstitial mechanism for 48g–96h′ migration path. The 96h′–O–96h′ migration path with the kick-out mechanism is delocalized and connected to the neighboring cells in three-dimensional space. Figures adapted with permission from Ref. [46].

Despite that $O_i$ cannot be easily formed under realistic condition in $Ca_3Fe_2Ge_3O_{12}$, high oxygen-ion conductivity was experimentally realized in donor-doped garnet. Doping with higher-valent cations can not only create more interstitial oxygens but also supply mobile electrons. Measurements on donor-doped samples detect both oxide-ion and electronic conduction.[47] In this work, they screened 90 garnet oxides and identified several with low oxygen-ion migration barriers, notably $Cd_3Sc_2Ge_3O_{12}$, $Ca_3Y_2Ge_3O_{12}$, $Ca_3In_2Ge_3O_{12}$, and $Ca_3Ga_2Ge_3O_{12}$ with barriers of 0.34-0.45 eV, which are expected to have oxygen ion conductivities of $\sim 10^{-2}$ S/cm at 700 °C. As research in this field progresses, garnet oxides are expected to play an increasingly important role in oxygen-active applications.

### 2.12 Perrierite/chevkinite $La_4Mn_5Si_4O_{22+\delta}$

Perrierite/chevkinite was first discovered as a new family of interstitial oxygen-ion conductor in 2024.[48] Through a combination of physically-motivated structure and property descriptors, *ab initio* simulations, and experiments, Meng *et al.* demonstrate an approach to discover new fast interstitial oxygen conductors out of ~34k oxide materials, utilizing a set of screening criteria including the geometric free space, thermodynamic stability, synthesizability, redox-active elements, and presence of short diffusion pathways.[48] $La_4Mn_5Si_4O_{22+\delta}$ (LMS), as a representative member of perrierite/chevkinite was identified as a new family of interstitial oxygen conductor. LMS, crystallized in the space group *C2/m*, features a layer structure consist layered sorosilicate and rutile-like sheets of edge-shared $Mn_1^{4+}/Mn_2^{3+}$ octahedra layers, as shown in **Figure 15**. Sorosilicate $Si_2O_7$ groups show a zigzag arrangement along the a-axis, connect with $Mn_3^{2+}$ octahedra along b-axis and $Mn_2^{3+}$ octahedra along c-axis by sharing corners, leaving free space in between these unconnected $Si_2O_7$ chains. The La atoms are between the rutile-like layer and the sorosilicate layer, surrounded by 10 oxygen atoms.

Meng *et al.* confirmed the composition of synthesized LMS as $La_{4.00}Mn_{4.69}Si_{4.03}O_{22.42}$, employed electron probe micro-analyzer (EPMA) analysis, suggesting significant excess oxygen with $\delta = 0.42$. Ionic conductivity of LMS was measured using 4-probe method with 8YSZ as electron blocking, reached to 0.05 S/cm at 800°C, comparable to many of the best fast oxygen conductors. *Ab initio* studies and simple thermodynamic considerations suggest that excess oxygen with $La_4Mn_5Si_4O_{22+\delta}$ ($\delta \approx 0.5$) is thermodynamically favorable under air conditions. The most stable interstitial site ($O_i^1$) lies in between two adjacent sorosilicate $Si_2O_7$ groups, connecting two Si tetrahedra. The second most stable interstitial site ($O_i^2$) lies in the joint of the $Si_2O_7$ and the $Mn_3^{2+}$ octahedra. These two prevailing interstitial sites contribute two distinct and competitive diffusion pathways. In the interstitial diffusion mechanism (yellow arrow in **Figure 15**), the $O_i$ hops between the $O_i^1$ sites through the channel between sorosilicate chains along the a-axis. A parallel active interstitialcy (cooperative “knock-on”) mechanism is indicated by cyan arrows, in which the $O_i$ moves along the corner-sharing $Si_2O_7$-$MnO_2$-$Si_2O_7$ framework along the b-axis. In the interstitialcy mechanism, the $O_i$ first hops from the $O_i^1$ site to a lattice site by kicking a lattice oxygen to the $O_i^2$ site, which then moves to a lattice site by kicking another lattice oxygen to the next $O_i^1$ site. NEB calculated energy landscape along the diffusion pathways reveals migration barriers of 0.69 eV and 0.74 eV for the interstitial and interstitialcy mechanisms, respectively. The high ionic conductivity in LMS is attributed to these two competitive diffusion paths. LMS, as a new family of interstitial oxygen conductors, demonstrates that its ample free volume, flexible corner-sharing polyhedral network, and redox-active $Mn_3^{2+}$ ions, make it highly suitable for the facile incorporation and transport of interstitial oxygen ions.

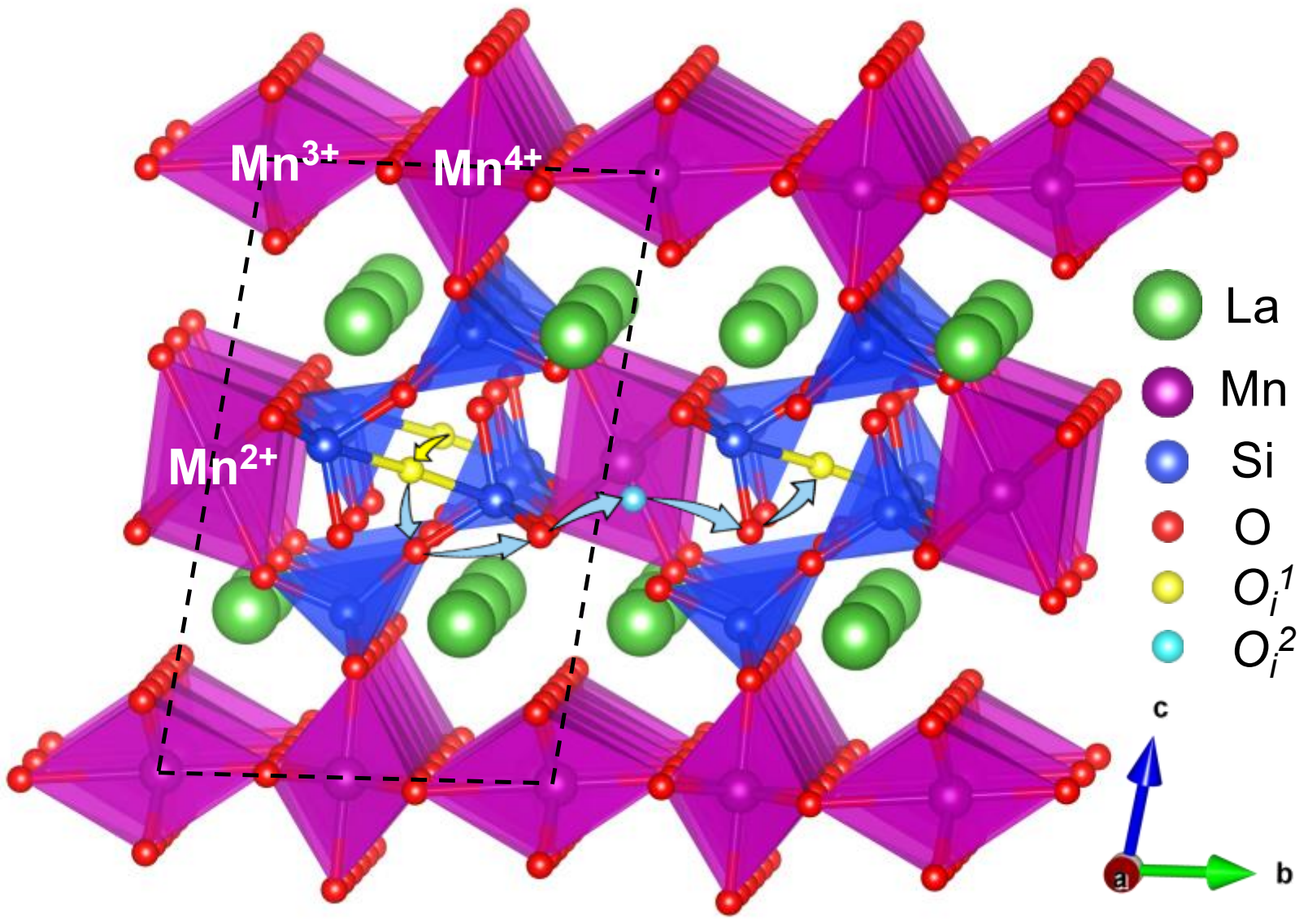

Figure 15 Bulk structure of $La_4Mn_5Si_4O_{22}$. The La, Mn, Si and O sites are shown as green, purple, blue and red spheres, respectively. The black dashed line denotes the single unit cell. The yellow ball represents the site and the cyan ball represents the site, respectively. Interstitial oxygen ($O_i$) in LMS diffuses through both interstitial mechanism (yellow arrow) and interstitialcy (cooperative 'knock-on') mechanism (cyan arrows). Figures adapted with permission from Ref. [48].

# 3 Design Strategies for Interstitial Oxygen Conductors

The design and discovery of new interstitial oxygen-ion conductors have evolved significantly, moving beyond traditional doping techniques to more sophisticated and systematic approaches. In this section, we discuss the design strategies employed to develop these materials, drawing from examining known families.

## 3.1 Donor doping

Donor doping remains one of the most widely applied strategies for enhancing interstitial oxygen-ion conductivity. By introducing aliovalent cations, donor doping supplies electrons, thereby lowering the energy cost for the oxygen reduction reaction and facilitating the formation of interstitial oxygen ions, thus increases the concentration of interstitial oxygens and boosts oxide-ion conductivity. It can also create additional free volume when smaller dopants replace larger host cations, and help stabilize the chemical and electronic environment, further enhancing the material's overall performance.

This approach has been successfully applied to nearly all known families of interstitial oxygen conductors. One notable example is that palmierite was confirmed as interstitial oxygen-ion conducting materials through targeted donor-doping. On the basis of studies of interstitial oxygen-ion conducting in the palmierite layer in hexagonal perovskites, Tawse *et al*. synthesized the palmierite $Ba_3Ti_{0.9}Mo_{1.1}O_{8.1}$ by targeted donor doping with $Mo^{6+}$ substitute for $Ti^{4+}$.[40] Neutron diffraction data verify the presence of interstitial oxygen ions in the $[BaO_{2+x}]$ layer, where oxygen transport proceeds via an interstitialcy mechanism. This resulted in a total ionic conductivity of 3.9 mS/cm at 600 °C, exceeding that of hexagonal perovskite. This example showcases that, with the diffusion pathways and structural features are well understood, donor-doping in materials with similar local structures is a straightforward and effective route to designing new high-performing interstitial oxygen conductors.

## 3.2 Element screening

Building on donor doping, element screening has become a more refined approach to identifying new interstitial oxygen-ion conducting materials. Researchers have screened extensive material databases by correlating migration barriers with structural features, such as ionic radii and bond lengths, alongside computational modeling (e.g., bond valence-based energy calculations, *ab* *initio* calculations) and experiments to assess oxygen-ion migration barriers and ionic conductivities. This screening process has successfully identified several new promising interstitial oxygen conductors.

Since that high transport number of the oxygen ion can be expected in Ga-containing oxides due to the $d^{10}$ electron configuration of the $Ga^{3+}$ and the reality practice that many gallates, such as LSGM, Melilite $La_{1.54}Sr_{0.46}Ga_3O_{7.27}$, $BaLaGaO_4$, $LaSrGaO_4$, and $Nb_3GaO_6$ have been reported to exhibit high oxygen-ion conductivities. Yasui *et al.* employed a screening strategy on 217 Ga-containing oxides, and identified the orthorhombic calcium gallate $Ca_3Ga_4O_9$, as a candidate of oxygen-ion conductor with a relatively low migration barrier and predicted ionic conductivity of $1.04\times10^{-5}$ S/cm at 800 °C.[152] $Ca_3Ga_4O_9$ is a layered crystal structure consisting of two-dimensional (2D) corner-sharing $GaO_4$ tetrahedral network, sharing similar structural and transport properties to melilite $LaSrGa_3O_7$. This discovery highlights the successful identification of a rare-earth-free oxygen-ion conductor with promising transport properties and demonstrates the effectiveness of low-cost bond valence analysis in discovering new functional materials.

In a parallel study, J. Lee *et al.* have screened 90 garnet-type oxides with high-throughput *ab initio* calculations and identified several candidates with low oxygen-ion migration barriers. The authors generated 90 combinations of $A_3B_2C_3O_{12}$ with typical valences of cations A, B, and C of $2^+$, $3^+$, and $4^+$, respectively. The selection of cations in the garnet crystal structure are shown in **Figure 16a**. It also shows the workflow of the screening in the search for energetically and dynamically stable garnet-type oxides with low migration barrier ($E_{mig}$) for the $O_i$ along the interstitialcy diffusion pathway (96h′-O-96h′). In condition 1, the formation energies of the garnet-type oxides with respect to their corresponding binary oxides were evaluated for its phase stability. Oxides with positive formation energy and small band gap were excluded, in which the band gap criterion is used to avoid the possibility of electronic conduction at high temperature. In condition 2, 16 oxides were retained with DFT NEB calculated $E_{mig}$ below 0.5 eV. Then, oxides with negative phonon frequencies were excluded in condition 3. Last, the computed formation energy was compared to the convex-hull states from Materials Project to focus on the most stable compounds. Among the final candidates, four compounds $Cd_3Sc_2Ge_3O_{12}$, $Ca_3Y_2Ge_3O_{12}$, $Ca_3In_2Ge_3O_{12}$, and $Ca_3Ga_2Ge_3O_{12}$ were found with barriers between 0.34 and 0.45 eV. These promising materials are predicted to achieve oxygen-ion conductivities around $10^{-2}$ S/cm at 700 °C, comparable to current high-performance electrolyte YSZ.[47]

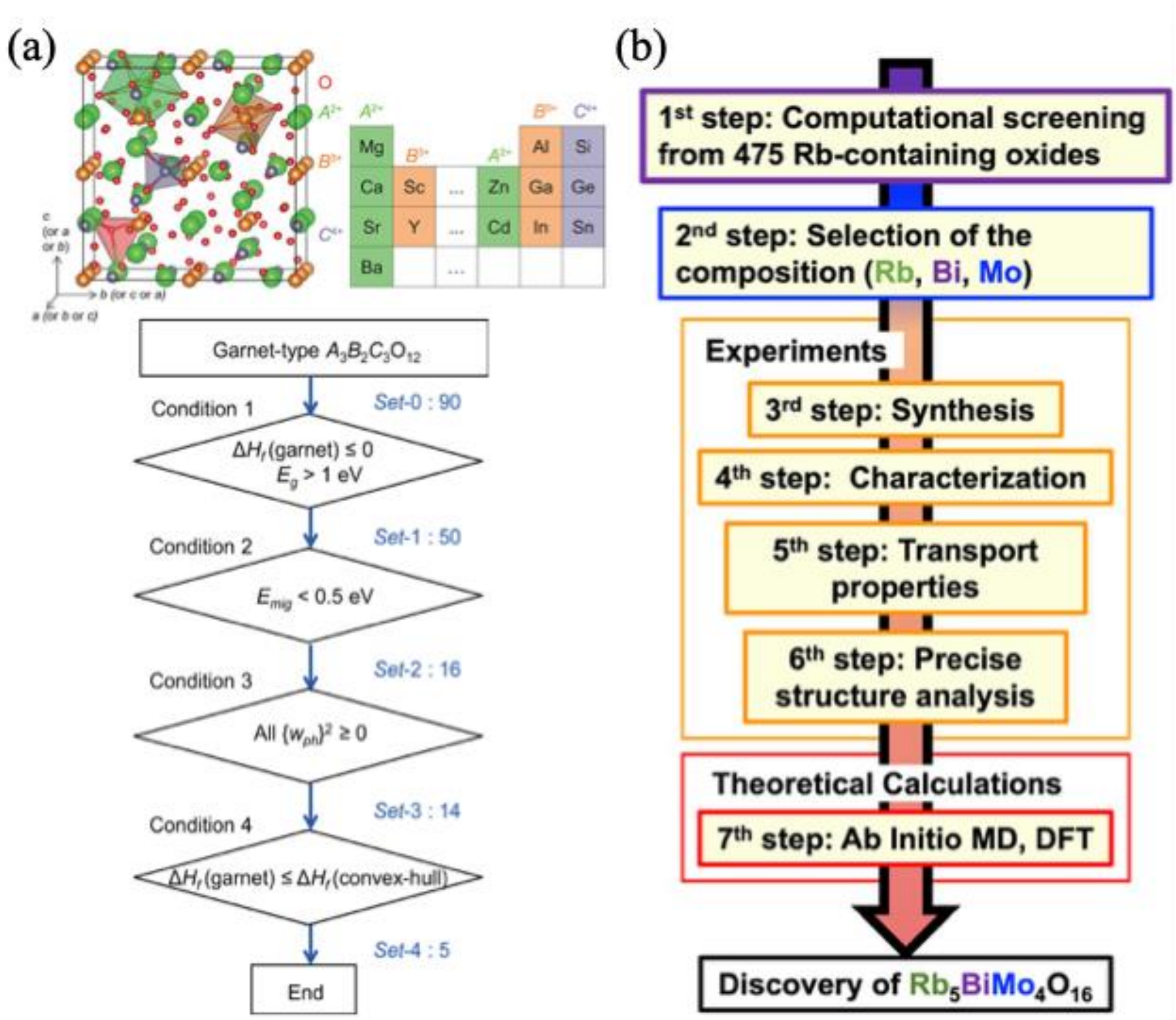

Figure 16 (a) Upper: Garnet crystal structures and cations selection used to generate 90 combinations of garnet-type $A_3B_2C_3O_{12}$ compositions. The numerical values above the periodic table are presumed valences of cations. Lower: Workflow for screening for garnet-type $A_3B_2C_3O_{12}$ with the screening conditions for stability and low $E_{mig}$. Parameters $\Delta H_f$, $E_g$, $E_{mig}$, and $w_{ph}$ denote the relative formation energy compared with binary oxides, band gap, migration barrier energy of $O_i$, and the eigenvalue of phonons, respectively. Condition 3 ensures dynamical stability. (b) Scheme of the computer screening of 475 Rb-containing oxides, composition selection, and experimental and theoretical demonstration of the Rb-containing high oxide-ion conductor $Rb_5BiMo_4O_{16}$. Figures adapted with permission from Ref. [47,153].

$Rb^+$ is the second largest available cation (after $Cs^+$), having it in an oxide structure can expand the lattice and increase free volume, thereby reducing the migration barrier for oxygen transport. Through a combined computational and experimental approach that particularly focused on Rb-based materials, as shown in **Figure 16b**, Yasui *et al*. screened 475 Rb-containing oxides and identified $Rb_5BiMo_4O_{16}$, which exhibited a high oxygen-ion conductivity of $2.3\times10^{-3}$ S/cm at 560 °C and $1.4\times10^{-4}$ S/cm at 300 °C, outperforming YSZ at lower temperatures.[153] The high ion mobility was attributed to the significant atomic motion in its $MoO_4$ tetrahedra and large structural free volume. Besides, the authors have shown similar high conductivities in other Rb-based palmierite oxides $Rb_5RMo_4O_{16}$, where R is a rare-earth cation, revealing new opportunities for designing low-temperature oxide-ion conductors by harnessing large-cation chemistry.

### 3.3 Physical-intuition descriptor and Machine Learning

A more advanced strategy involves a descriptor-guided materials discovery approach informed by physical intuition, where the descriptors serve as criteria to assist in materials screening for a broader materials dataset. Meng *et al*.[48] designed an effective screening approach based on physical intuition derived structural and chemical features (free volume, short hop distance, thermodynamic stability, oxidizability, and synthesizability), *ab initio* calculations, and experiments to search for new families of interstitial oxygen diffusers from many cataloged oxides, as the screening scheme shown in **Figure 17a**. This descriptor screening step winnowed the field of 34k oxides down to just 345 oxides, a 99% reduction in the search space considered by basic descriptor analysis of material structure and composition. Three new families, comprising the molybdates $A_2TM_2(MoO_4)_3$ (A=alkali metal, TM=transition metal), perrierite/chevkinite $RE_4TM_5Si_4O_{22}$ (RE=rare earth, TM=transition metal), and germinates $RE_1TM_2Ge_4O_{12}$ (RE= rare earth, TM=transition metal), were predicted as promising interstitial oxygen conductors and exhibit completely different structure types from any known fast oxygen conductors. Detailed studies on $La_4Mn_5Si_4O_{22+\delta}$ (LMS), confirmed it as a new interstitial oxygen conductor. *Ab initio* and experimental results show that LMS has higher oxygen ionic conductivity compared to the YSZ and one of highest oxygen surface exchanges of known materials at lower temperatures. The success of this approach suggests that these relatively simple structural and chemical descriptors are strongly correlated with stable interstitial oxygen formation and fast migration, suggesting that many more materials exhibiting fast interstitial oxygen kinetics remain to be found and that this approach developed in this work is a practical path to their discovery.

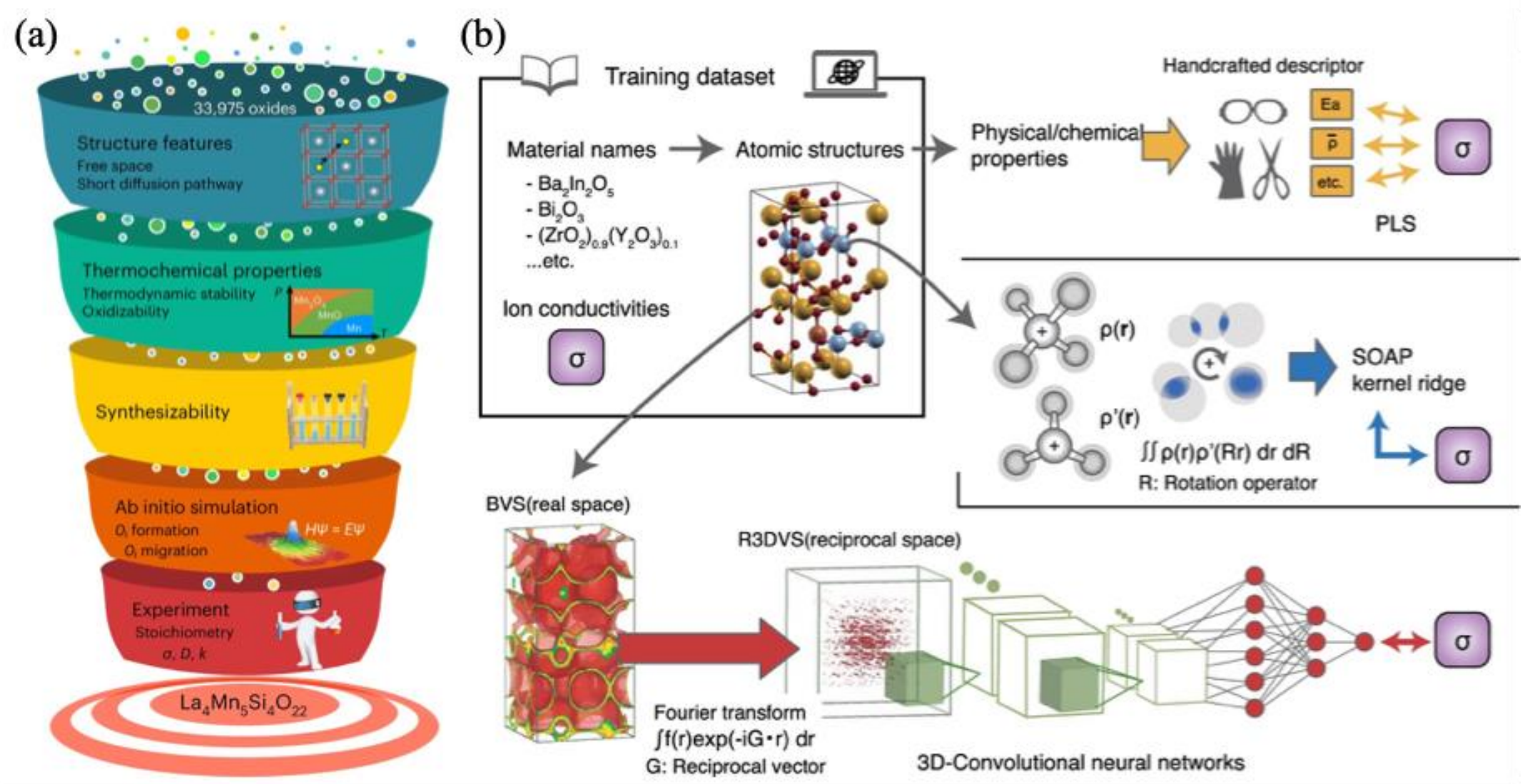

Figure 17 (a) High-throughput screening approach for discovering interstitial oxygen conductors using a set of physical intuition derived structural and chemical features (free volume, short hop distance, thermodynamic stability, oxidizability, and synthesizability), *ab initio* calculations, and experiments. (b) Schematic picture composed of three independent machine learning models for the oxygen-ion conductivity as the target property. The first model employs a knowledge-based handcrafted descriptor whose dimensionality is optimally reduced by the PLS model. The second descriptor is SOAP, which calculates a metric between compounds with respect to the atomic geometry and electronic negativity. The SOAP is used for the kernel-ridge regression model. The third descriptor is R3DVS. This descriptor encodes the BVS voxel data, which approximate the oxygen-ion transportation path into a fixed-length array that is suitable for the 3D-CNN model.

In another case, Kajita *et al*. developed an "ensemble-scope descriptor" approach using only 29 known materials as training data, to screen for new oxygen-ion conductors among more than 13,000 candidate oxides. Three distinct descriptors: (1) a handcrafted descriptor emphasizing chemical and structural features, (2) a SOAP (smooth overlap of atomic positions) descriptor capturing local atomic environments, and (3) a R3DVS (reciprocal 3D voxel space) descriptor encoding the bond-valence distribution are used to train three separate machine learning models. As the schematic picture shown in **Figure 17b**, three machine learning models were trained independently to predict the ionic conductivity $\sigma$ with the dataset of 29 oxygen-ion conductors, with one handcrafted and two generic descriptors to maximize the inference ability even using a small training dataset. By averaging their predictions, the authors overcame the usual "small data" issue in materials discovery and identified promising compounds rapidly. 18 oxides with a predicted conductivity higher than $10^{-4}$ S/cm were selected for experimental validation. Eight oxides including $EuGe_2KO_6$, $Ca_3Fe_2Ge_3O_{12}$, $BaCu_2Ge_2O_7$, $DyAlO_3$, and four bismuth oxides were found as oxygen-ion conductors, in which five new oxygen-ion conductors lie well outside classical conductor families. This work showcased that effective machine learning model with a small set of training data can expedite the materials discovery guided by carefully chosen descriptors.[150]

These evolving design strategies, from donor doping and element screening to knowledge driven descriptor-guided materials discovery, highlight a powerful shift toward systematic and predictive design of interstitial oxygen-ion conductors. By integrating fundamental understanding of conduction mechanisms with computational screening and experimental validation, researchers are rapidly uncovering new materials with high performance across a range of operating temperatures. As these approaches continue to mature, the discovery of novel interstitial oxygen conductors is expected to accelerate, offering promising pathways for next-generation energy technologies such as solid oxide fuel cells, oxygen separation membranes, and electrochemical reactors.

# 4 Advancing the Discovery of Interstitial Oxygen Conductors

In contrast to vacancy oxygen conductors, interstitial oxygen conductors have historically received far less attention, and until recently, little was known about how to rationally design materials with fast interstitial oxygen transport. In the extensive studies on vacancy-mediated oxygen conductors, several design principles have been proposed. These include the idea of "partial occupancy of energetically equivalent oxide ion lattice sites", as seen in fluorite structures, and the strategy of "partial substitution of oxygen for vacancies occupied by lone pairs", as exemplified by $La_2Mo_2O_9$ materials.[16] By comparison, the design space for interstitial oxygen conductors has remained largely underexplored. Recent advances, however, have begun to illuminate key features that govern interstitial conduction. Through a comprehensive review of known interstitial oxygen-ion conducting materials, we identified that high-performing interstitial oxygen-ion conductors consistently exhibit two important features: (1) the availability of electrons, typically achieved through oxidizable cations or aliovalent doping, and (2) structural flexibility with accessible volume, often provided by low-connectivity polyhedral frameworks, are critical for incorporating interstitial oxygen-ion and enabling fast ion conduction.[48]

In light of these insights, we propose a knowledge-driven approach to identify promising interstitial oxygen conductors that goes beyond conventional trial-and-error or structure-type-based searches. Specifically, materials that combine two critical traits: readily available electrons and structural flexibility, emerge as prime candidates for high-performance interstitial conduction. We outline three practical and scalable strategies to guide this search: (1) Systematic screening for materials featuring a corner-sharing polyhedral network or isolated polyhedral combined with redox-active cations capable of donating electrons; (2) Cation substitution engineering to supply electrons for materials with corner-sharing polyhedral or isolated polyhedral, thereby promoting interstitial oxygen formation and mobility; and (3) Exploration of alternative structural motifs with high structural flexibility and accessible free volume, which facilitate ion incorporation and diffusion. We propose promising structural motifs for this exploration: layered materials, soft lattice materials, materials with low polyhedral connectivity, and materials featuring a mixture of large cations (alkali, alkali-earth, rare-earth families) and small cations (transition metal, post-transition metal).

The rapid growth of high-quality materials databases, combined with the continued advancement of data-centric machine learning (ML) approaches, has opened up unprecedented opportunities for discovering new interstitial oxygen-ion conductors. While datasets specifically for interstitial oxygen conduction remain limited, supervised and semi-supervised ML techniques show significant potential in accelerating the identification of these materials. Similar approaches have successfully been applied to the discovery of novel Li-ion conducting materials and oxygen vacancy-mediated conductors, whereas interstitial oxygen-ion conductors remain less explored. These successes highlight the untapped potential of ML-guided screening to uncover new families of interstitial oxygen-ion conductors at scale. By integrating fundamental design principles, materials informatics, and computational validation, the discovery of high-performance interstitial conductors can be greatly accelerated.

# 5 Summary

This review provided a comprehensive summary of the current state of research on interstitial oxygen conductors. Unlike traditional vacancy-mediated oxygen conductors, interstitial oxygen conductors offer significantly lower migration barriers, making them particularly efficient at lower operating temperatures (400-600°C). This review systematically examines 12 families of interstitial oxygen conductors, organized by their key structural features critical for interstitial incorporation and diffusion: (i) Fluorite $UO_{2+x}$ and sillén oxychloride, with highly symmetric oxygen sublattices and well-defined interstitial sites facilitating fast interstitialcy diffusion; (ii) Scheelite, apatite, hexagonal perovskite derivatives and palmierite, with isolated tetrahedra that provide local structural flexibility and open pathways for ion migration; (iii) Ruddlesden-Popper perovskite, melilite, and hexagonal manganite, with two-dimensional corner-sharing polyhedral networks that accommodate planar diffusion pathways; (iv) Mayenite, langasite, and garnet, with three-dimensional corner-sharing polyhedral frameworks offering interconnected diffusion channels; and (v) Perrierite/chevkinite $La_4Mn_5Si_4O_{22+\delta}$, with a combination of isolated sorosilicate units and corner-sharing networks which enabled both interstitial and interstitialcy diffusing mechanisms, emphasizing their structural characteristics, ionic conductivities, and diffusion mechanisms of interstitial oxygen-ions. Beyond summarizing known materials, this review discussed the universal structural features that enable fast oxygen-ion conduction and outlined recent key advancements in computational and experimental approaches to discovering new interstitial oxygen conductors. These approaches have evolved from donor doping and element screening to more sophisticated, physical-intuition descriptor-guided screening and machine learning. This review highlighted two critical factors governing the performance of interstitial oxygen-ion conductors: (1) the availability of electrons, typically facilitated by oxidizable cations or aliovalent doping, and (2) the structural flexibility with accessible volume of the material, both of which are essential for accommodating and enabling the rapid migration of interstitial oxygen ions. Finally, a knowledge-driven methodology that leverages current understanding to identify promising interstitial oxygen conductors outside traditional search paradigms is proposed for the future directions in interstitial oxygen conducting material discovery and design. In addition, the review discussed the emerging potential of machine learning (ML) and data-centric approaches in accelerating the discovery process. With the aid of large-scale materials databases and advanced computational tools, ML-guided screening is expected to significantly reduce development timelines and identify unexplored material classes. Together, these advances pave the way for the next generation of high-performance interstitial oxygen-ion conductors, with broad applications in solid oxide fuel cells, oxygen separation membranes, and other oxygen-active technologies essential for a sustainable energy future.

**Data availability**

Data sharing is not applicable to this article as no new data were created or analyzed in this study.

**CRediT authorship contribution statement**

Jun Meng: Conceptualization, Writing – review & editing.

**Declaration of Competing Interest**

The authors declare that they have no known competing financial interests or personal relationships that could have appeared to influence the work reported in this paper.

**Acknowledgement**

Support for J.M. was provided by US Department of Energy (DOE), Office of Science, Basic Energy Sciences (BES), under Award No. DE-SC0020419. Additional support for J.M. was the Advanced Cyberinfrastructure Coordination Ecosystem: Services & Support (ACCESS) program supported by National Science Foundation (NSF) grants #2138259, #2138286, #2138307, #2137603, and #2138296.

J.M thanks Dane Morgan and Ryan Jacobs for helpful discussions in preparing this review.

**References**

[1] R. Zohourian, R. Merkle, G. Raimondi, J. Maier, Mixed-Conducting Perovskites as Cathode Materials for Protonic Ceramic Fuel Cells: Understanding the Trends in Proton Uptake, Adv Funct Mater 28 (2018) 1801241. https://doi.org/10.1002/adfm.201801241.

[2] G. Kobayashi, Y. Hinuma, S. Matsuoka, A. Watanabe, M. Iqbal, M. Hirayama, M. Yonemura, T. Kamiyama, I. Tanaka, R. Kanno, Pure H– conduction in oxyhydrides, Science (1979) 351 (2016) 1314–1317. https://doi.org/10.1126/science.aac9185.

[3] E.D. Wachsman, K.T. Lee, Lowering the temperature of solid oxide fuel cells, Science (1979) 334 (2011) 935–939. https://doi.org/10.1126/science.1204090.

[4] Y. Zhang, R. Knibbe, J. Sunarso, Y. Zhong, W. Zhou, Z. Shao, Z. Zhu, Recent Progress on Advanced Materials for Solid-Oxide Fuel Cells Operating Below 500 °C, Advanced Materials 29 (2017) 1700132. https://doi.org/10.1002/adma.201700132.

[5] S.H. Jensen, P.H. Larsen, M. Mogensen, Hydrogen and synthetic fuel production from renewable energy sources, Int J Hydrogen Energy 32 (2007) 3253–3257. https://doi.org/https://doi.org/10.1016/j.ijhydene.2007.04.042.

[6] A. Manthiram, X. Yu, S. Wang, Lithium battery chemistries enabled by solid-state electrolytes, Nat Rev Mater 2 (2017) 1–16. https://doi.org/10.1038/natrevmats.2016.103.

[7] J. Janek, W.G. Zeier, Challenges in speeding up solid-state battery development, Nat Energy 8 (2023) 230–240. https://doi.org/10.1038/s41560-023-01208-9.

[8] C. Zhang, K. Huang, Solid-oxide metal–air redox batteries, in: Solid Oxide-Based Electrochemical Devices, Elsevier, 2020: pp. 217–250. https://doi.org/10.1016/B978-0-12-818285-7.00007-1.

[9] T. Famprikis, P. Canepa, J.A. Dawson, M.S. Islam, C. Masquelier, Fundamentals of inorganic solid-state electrolytes for batteries, Nat Mater 18 (2019) 1278–1291. https://doi.org/10.1038/s41563-019-0431-3.

[10] K. Shan, Z. Yi, J. Wang, Limiting current oxygen sensor based on Y, In co-doped SrTiO3 as a dense diffusion barrier layer, Nano Res 12 (2021) 3–8. https://doi.org/10.1007/s12274-021-3379-y.

[11] X. Gao, T. Liu, J. Yu, L. Li, Limiting current oxygen sensor based on La0.8Sr0.2Ga0.8Mg0.2O3−δ as both dense diffusion barrier and solid electrolyte, Ceram Int 43 (2017) 6329–6332. https://doi.org/10.1016/j.ceramint.2017.02.040.

[12] X. Zhu, W. Yang, Microstructural and Interfacial Designs of Oxygen-Permeable Membranes for Oxygen Separation and Reaction–Separation Coupling, Advanced Materials 31 (2019) 1902547. https://doi.org/10.1002/adma.201902547.

[13] J. Adánez, L.F. De Diego, F. García-Labiano, P. Gayán, A. Abad, J.M. Palacios, Selection of oxygen carriers for chemical-looping combustion, Energy and Fuels 18 (2004) 371–377. https://doi.org/10.1021/ef0301452.

[14] V. Goldstein, M.K. Rath, A. Kossenko, N. Litvak, A. Kalashnikov, M. Zinigrad, Solid oxide fuel cells for ammonia synthesis and energy conversion, Sustain Energy Fuels 6 (2022) 4706–4715. https://doi.org/10.1039/D2SE00954D.

[15] S.D. Günay, Investigation of Oxygen Diffusion in Irradiated UO2 with MD Simulation, High Temperature Materials and Processes 35 (2016) 981–987. https://doi.org/10.1515/htmp-2015-0137.

[16] J.B. Goodenough, Oxide-ion conductors by design, Nature 404 (2000) 821–823. https://doi.org/10.1038/35009177.

[17] B.C.H. Steele, Appraisal of Ce1-yGdyO2-y/2 electrolytes for IT-SOFC operation at 500 °C, Solid State Ion 129 (2000) 95–110. https://doi.org/10.1016/S0167-2738(99)00319-7.

[18] R.D. Bayliss, S.N. Cook, S. Kotsantonis, R.J. Chater, J.A. Kilner, Oxygen ion diffusion and surface exchange properties of the α- And δ-phases of Bi2O3, Adv Energy Mater 4 (2014) 1301575. https://doi.org/10.1002/aenm.201301575.

[19] J.A. Díaz-Guillén, M.R. Díaz-Guillén, K.P. Padmasree, A.F. Fuentes, J. Santamaría, C. León, High ionic conductivity in the pyrochlore-type Gd2−yLayZr2O7 solid solution (0≤y≤1), Solid State Ion 179 (2008) 2160–2164. https://doi.org/https://doi.org/10.1016/j.ssi.2008.07.015.

[20] J. Drennan, V. Zelizko, D. Hay, F.T. Ciacchi, S. Rajendran, S. P. S. Badwal, Characterisation, conductivity and mechanical properties of the oxygen-ion conductor La0.9Sr0.1Ga0.8 Mg0.2O3-x, J Mater Chem 7 (1997) 79–83. https://doi.org/10.1039/A604563D.

[21] G.M. Rupp, M. Glowacki, J. Fleig, Electronic and Ionic Conductivity of La0.95Sr0.05Ga0.95Mg0.05O3-δ (LSGM) Single Crystals, J Electrochem Soc 163 (2016) F1189. https://doi.org/10.1149/2.0591610jes.

[22] A. Fluri, E. Gilardi, M. Karlsson, V. Roddatis, M. Bettinelli, I.E. Castelli, T. Lippert, D. Pergolesi, Anisotropic Proton and Oxygen Ion Conductivity in Epitaxial Ba2In2O5 Thin Films, Journal of Physical Chemistry C 121 (2017) 21797–21805.

https://doi.org/10.1021/acs.jpcc.7b02497.

[23] S. Xu, R. Jacobs, D. Morgan, Factors Controlling Oxygen Interstitial Diffusion in the Ruddlesden–Popper Oxide La2–xSrxNiO4+δ, Chemistry of Materials 30 (2018) 7166–7177. https://doi.org/10.1021/acs.chemmater.8b03146.

[24] W. Zhang, K. Fujii, E. Niwa, M. Hagihala, T. Kamiyama, M. Yashima, Oxide-ion conduction in the Dion–Jacobson phase CsBi2Ti2NbO10−δ, Nat Commun 11 (2020) 1–8. https://doi.org/10.1038/s41467-020-15043-z.

[25] K.R. Kendall, C. Navas, J.K. Thomas, H.C. zur Loye, Recent Developments in Oxide Ion Conductors: Aurivillius Phases, Chem. Mater. 8 (1996) 642–649. https://doi.org/10.1016/0167-2738(95)00207-4.

[26] M. Kluczny, J.T. Song, T. Akbay, E. Niwa, A. Takagaki, T. Ishihara, Sillén-Aurivillius phase bismuth niobium oxychloride, Bi4NbO8Cl, as a new oxide-ion conductor, J Mater Chem A Mater 10 (2022) 2550–2558. https://doi.org/10.1039/d1ta07335d.

[27] Y. Zhu, W. Zhou, Y. Chen, Z. Shao, An Aurivillius Oxide Based Cathode with Excellent CO2Tolerance for Intermediate-Temperature Solid Oxide Fuel Cells, Angewandte Chemie - International Edition 55 (2016) 8988–8993. https://doi.org/10.1002/anie.201604160.

[28] F. Abraham, J.C. Boivin, G. Mairesse, G. Nowogrocki, The bimevox series: A new family of high performances oxide ion conductors, Solid State Ion 40–41 (1990) 934–937. https://doi.org/10.1016/0167-2738(90)90157-M.

[29] P. Lacorre, F. Goutenoire, O. Bohnke, R. Retoux, Y. Laligant, Designing fast oxide-ion conductors, Nature 404 (2000) 9–11.

[30] J. Belle, Oxygen and Uranium diffusion in Uranium dioxide, Journal of Nuclear Materials 30 (1969) 3–15.

[31] B. Dorado, P. Garcia, G. Carlot, C. Davoisne, M. Fraczkiewicz, B. Pasquet, M. Freyss, C. Valot, G. Baldinozzi, D. Siméone, M. Bertolus, First-principles calculation and experimental study of oxygen diffusion in uranium dioxide, Phys Rev B Condens Matter Mater Phys 83 (2011) 1–10. https://doi.org/10.1103/PhysRevB.83.035126.

[32] J. Meng, M.S. Sheikh, L.E. Schultz, W.O. Nachlas, J. Liu, M.P. Polak, R. Jacobs, D. Morgan, Ultrafast Oxygen Conduction in Sillén Oxychlorides, Chemistry of Materials 37 (2025) 3147–3157. https://doi.org/10.1021/acs.chemmater.4c03163.

[33] H. Yaguchi, D. Morikawa, T. Saito, K. Tsuda, M. Yashima, High Oxide-Ion Conductivity through the Interstitial Oxygen Site in Sillén Oxychlorides, Adv Funct Mater 33 (2023) 2214082. https://doi.org/10.1002/adfm.202214082.

[34] N. Ueno, H. Yaguchi, K. Fujii, M. Yashima, High Conductivity and Diffusion Mechanism of Oxide Ions in Triple Fluorite-Like Layers of Oxyhalides, J Am Chem Soc 146 (2024) 11235–11244. https://doi.org/10.1021/jacs.4c00265.

[35] J. Li, F. Pan, S. Geng, C. Lin, L. Palatinus, M. Allix, X. Kuang, J. Lin, J. Sun, Modulated structure determination and ion transport mechanism of oxide-ion conductor CeNbO4+δ, Nat Commun 11 (2020) 1–9. https://doi.org/10.1038/s41467-020-18481-x.

[36] R.J. Packer, S.J. Skinner, A.A. Yaremchenko, E. V. Tsipis, V. V. Kharton, M. V. Patrakeev, Y.A. Bakhteeva, Lanthanum substituted CeNbO4+δ scheelites: Mixed conductivity and structure at elevated temperatures, J Mater Chem 16 (2006) 3503–3511. https://doi.org/10.1039/b606261j.

[37] H. Arikawa, H. Nishiguchi, T. Ishihara, Y. Takita, Oxide ion conductivity in Sr-doped La10Ge6O27 apatite oxide, Solid State Ion 136–137 (2000) 31–37. https://doi.org/10.1016/S0167-2738(00)00386-6.

[38] M. Yashima, T. Tsujiguchi, Y. Sakuda, Y. Yasui, Y. Zhou, K. Fujii, S. Torii, T. Kamiyama, S.J. Skinner, High oxide-ion conductivity through the interstitial oxygen site in Ba7Nb4MoO20-based hexagonal perovskite related oxides, Nat Commun 12 (2021) 1–7. https://doi.org/10.1038/s41467-020-20859-w.

[39] S. Fop, K.S. McCombie, E.J. Wildman, J.M.S. Skakle, J.T.S. Irvine, P.A. Connor, C. Savaniu, C. Ritter, A.C. Mclaughlin, High oxide ion and proton conductivity in a disordered hexagonal perovskite, Nat Mater 19 (2020) 752–757. https://doi.org/10.1038/s41563-020-0629-4.

[40] D.N. Tawse, S. Fop, J.W. Still, O.J.B. Ballantyne, C. Ritter, Y. Zhou, J.A. Dawson, A.C. Mclaughlin, Unlocking the Potential of Palmierite Oxides: High Oxide Ion Conductivity via Induced Interstitial Defects, (2025). https://doi.org/10.1021/jacs.4c17849.

[41] J. Song, D. Ning, B. Boukamp, J.M. Bassat, H.J.M. Bouwmeester, Structure, electrical conductivity and oxygen transport properties of Ruddlesden-Popper phases Ln: N +1NinO3 n +1(Ln = La, Pr and Nd; N = 1, 2 and 3), J Mater Chem A Mater 8 (2020) 22206–22221. https://doi.org/10.1039/d0ta06731h.

[42] X. Kuang, M.A. Green, H. Niu, P. Zajdel, C. Dickinson, J.B. Claridge, L. Jantsky, M.J. Rosseinsky, Interstitial oxide ion conductivity in the layered tetrahedral network melilite structure, Nat Mater 7 (2008) 498–504. https://doi.org/10.1038/nmat2201.

[43] S.H. Skjærvø, E.T. Wefring, S.K. Nesdal, N.H. Gaukås, G.H. Olsen, J. Glaum, T. Tybell, S.M. Selbach, Interstitial oxygen as a source of p-type conductivity in hexagonal manganites, Nat Commun 7 (2016) 7491. https://doi.org/10.1038/ncomms13745.

[44] M. Lacerda; J. T. S. Irvine; F. P. Glasser; and A. R. West, High oxide ion conductivity in Ca12Al14O33, Nature 332 (1988) 525–526.

[45] M. Diaz-Lopez, J.F. Shin, M. Li, M.S. Dyer, M.J. Pitcher, J.B. Claridge, F. Blanc, M.J. Rosseinsky, Interstitial Oxide Ion Conductivity in the Langasite Structure: Carrier Trapping by Formation of (Ga,Ge)2O8 Units in La3Ga5- xGe1+ xO14+ x/2 ($0 < x \leq 1.5$), Chemistry of Materials 31 (2019) 5742–5758. https://doi.org/10.1021/acs.chemmater.9b01734.

[46] J. Lee, N. Ohba, R. Asahi, Oxygen conduction mechanism in Ca 3 Fe 2 Ge 3 O 12 garnet-type oxide, Sci Rep 9 (2019) 1–8. https://doi.org/10.1038/s41598-019-39288-x.

[47] J. Lee, N. Ohba, R. Asahi, Design rules for high oxygen-ion conductivity in garnet-type oxides, Chemistry of Materials 32 (2020) 1358–1370. https://doi.org/10.1021/acs.chemmater.9b02044.

[48] J. Meng, M.S. Sheikh, R. Jacobs, J. Liu, W.O. Nachlas, X. Li, D. Morgan, Computational discovery of fast interstitial oxygen conductors, Nat Mater 23 (2024) 1252–1258. https://doi.org/10.1038/s41563-024-01919-8.

[49] Citrination, (n.d.).

[50] J.W. Wang, R.C. Ewing, U. Becker, Average structure and local configuration of excess oxygen in UO 2+x, Sci Rep 4 (2014) 3–5. https://doi.org/10.1038/srep04216.

[51] X.M. Bai, A. El-Azab, J. Yu, T.R. Allen, Migration mechanisms of oxygen interstitial clusters in UO2, Journal of Physics Condensed Matter 25 (2013). https://doi.org/10.1088/0953-8984/25/1/015003.

[52] P. Garcia, M. Fraczkiewicz, C. Davoisne, G. Carlot, B. Pasquet, G. Baldinozzi, D. Siméone, C. Petot, Oxygen diffusion in relation to p-type doping in uranium dioxide, Journal of Nuclear Materials 400 (2010) 112–118. https://doi.org/https://doi.org/10.1016/j.jnucmat.2010.02.019.

[53] G.E. Murch, C.R.A. Catlow, Oxygen diffusion in UO2, ThO2 and PuO2. A review, Journal of the Chemical Society, Faraday Transactions 2: Molecular and Chemical Physics 83 (1987) 1157–1169. https://doi.org/10.1039/F29878301157.

[54] J. Meng, L. Schultz, R. Jacobs, D. Morgan, Discovery of New Fast Oxygen Conductors: Bi2MO4x (M= rare earth, X= halogen) Via Unsupervised Machine Learning, ECS Meeting Abstracts MA2023-01 (2023) 2783. https://doi.org/10.1149/MA2023-01402783mtgabs.

[55] A.W. Sleight, Crystal Chemistry and Catalytic Properties of Oxides with the Scheelite Structure, ACADEMIC PRESS, INC., 1977. https://doi.org/10.1016/b978-0-12-147450-8.50011-3.

[56] W. Van Loo, Crystal growth and electrical conduction of PbMoO4 and PbWO4, J Solid State Chem 14 (1975) 359–365. https://doi.org/10.1016/0022-4596(75)90056-0.

[57] T. Lu; B. C. H. Steele, Electrical conductivity of polycrystalline BiVO4 samples having the scheelite structure, 21 (1986) 339–342.

[58] R.J. Cava, R.S. Roth, T. Negas, H.S. Parker, D.B. Minor, Crystal chemistry, modulated structure, and electrical conductivity in the oxygen excess scheelite-based compounds La1−xThxNbO4+x2 and LaNb1−xWxO4+x2, J Solid State Chem 40 (1981) 318–329. https://doi.org/https://doi.org/10.1016/0022-4596(81)90398-4.

[59] C. Li, R.D. Bayliss, S.J. Skinner, Crystal structure and potential interstitial oxide ion conductivity of LnNbO4 and LnNb0.92W0.08O4.04 (Ln=La, Pr, Nd), Solid State Ion 262 (2014) 530–535. https://doi.org/https://doi.org/10.1016/j.ssi.2013.12.023.

[60] L. Hoffart, U. Heider, L. Jörissen, R.A. Huggins, W. Witschel, Transport properties of materials with the scheelite structure, Solid State Ion 72 (1994) 195–198. https://doi.org/10.1016/0167-2738(94)90146-5.

[61] T. Esaka, T. Mina-ai, H. Iwahara, Oxide ion conduction in the solid solution based on the scheelite-type oxide PbWO4, Solid State Ion 57 (1992) 319–325. https://doi.org/https://doi.org/10.1016/0167-2738(92)90165-L.
[62] S.S. Pramana, T. Baikie, T. An, M.G. Tucker, J. Wu, M.K. Schreyer, F. Wei, R.D. Bayliss, C.L. Kloc, T.J. White, A.P. Horsfield, S.J. Skinner, Correlation of Local Structure and Diffusion Pathways in the Modulated Anisotropic Oxide Ion Conductor CeNbO4.25, J Am Chem Soc 138 (2016) 1273–1279. https://doi.org/10.1021/jacs.5b11373.
[63] K. Toyoura, Y. Sakakibara, T. Yokoi, A. Nakamura, K. Matsunaga, Oxide-ion conduction via interstitials in scheelite-type LaNbO4: a first-principles study, J Mater Chem A Mater 6 (2018) 12004–12011. https://doi.org/10.1039/C8TA02859A.
[64] J. Wang, L. Zhou, Y. Wang, J. Xu, X. Yang, X. Kuang, Molecular dynamic simulation of interstitial oxide ion migration in Pb1-xLaxWO4+x/2 scheelite, J Solid State Chem 268 (2018) 16–21. https://doi.org/10.1016/j.jssc.2018.08.023.
[65] S. Nakayama, T. Kageyama, H. Aono, Y. Sadaokac, Ionic Conductivity of Lanthanoid Silicates, Ln10(SiO4)6O3 (Ln=La, Nd, Sm, Gd, Dy, Y, Ho, Er and Yb), J Mater Chem 5 (1995) 1801–1805.
[66] S. Nakayama, M. Sakamoto, M. Highchi, K. Kodaira, Ionic conductivities of apatite type NdX(SiO4)6O1.5X-12 (X = 9.20 and 9.33) single crystals, J Mater Sci Lett 19 (2000) 91–93. https://doi.org/10.1023/A:1006674708833.
[67] J.E.H. Sansom, D. Richings, P.R. Slater, Powder neutron diffraction study of the oxide-ion-conducting apatite-type phases, La9.33Si6O26 and La8Sr2Si6O26, Solid State Ion 139 (2001) 205–210. https://doi.org/10.1016/S0167-2738(00)00835-3.
[68] L. León-Reina, E.R. Losilla, M. Martínez-Lara, S. Bruque, M.A.G. Aranda, Interstitial oxygen conduction in lanthanum oxy-apatite electrolytes, J Mater Chem 14 (2004) 1142–1149. https://doi.org/10.1039/b315257j.
[69] J.R. Tolchard, M.S. Islam, P.R. Slater, Defect chemistry and oxygen ion migration in the apatite-type materials La9.33Si6O26 and La8Sr 2Si6O26, J Mater Chem 13 (2003) 1956–1961. https://doi.org/10.1039/b302748c.
[70] M.S. Islam, J.R. Tolchard, P.R. Slater, An apatite for fast oxide ion conduction, Chemical Communications 3 (2003) 1486–1487. https://doi.org/10.1039/b301179h.
[71] T.K. Schultze, J.P. Arnold, S. Grieshammer, Ab Initio Investigation of Migration Mechanisms in la Apatites, ACS Appl Energy Mater 2 (2019) 4708–4717. https://doi.org/10.1021/acsaem.9b00226.
[72] K. Matsunaga, Ionic conduction mechanisms of apatite-type lanthanum silicate and germanate from first principles, Journal of the Ceramic Society of Japan 125 (2017) 670–676. https://doi.org/10.2109/jcersj2.17097.
[73] J.R. Peet, A. Piovano, M.R. Johnson, I.R. Evans, Location and orientation of lone pairs in apatite-type materials: A computational study, Dalton Transactions 46 (2017) 15996–15999. https://doi.org/10.1039/c7dt03956e.
[74] K. Imaizumi, K. Toyoura, A. Nakamura, K. Matsunaga, Cooperative oxide-ion conduction in apatite-type lanthanum germanate-A first principles study, Journal of the Ceramic Society of Japan 125 (2017) 105–111. https://doi.org/10.2109/jcersj2.16260.
[75] E. Kendrick, M.S. Islam, P.R. Slater, Developing apatites for solid oxide fuel cells: Insight into structural, transport and doping properties, J Mater Chem 17 (2007) 3104–3111. https://doi.org/10.1039/b704426g.
[76] E. Kendrick, M.S. Islam, P.R. Slater, Atomic-scale mechanistic features of oxide ion conduction in apatite-type germanates, Chemical Communications (2008) 715–717. https://doi.org/10.1039/B716814D.
[77] M. Yashima, Diffusion pathway of mobile ions and crystal structure of ionic and mixed conductors-A brief review, Journal of the Ceramic Society of Japan 117 (2009) 1055–1059. https://doi.org/10.2109/jcersj2.117.1055.
[78] E. Béchade, O. Masson, T. Iwata, I. Julien, K. Fukuda, P. Thomas, E. Champion, Diffusion Path and Conduction Mechanism of Oxide Ions in Apatite-Type Lanthanum Silicates, Chemistry of Materials 21 (2009) 2508–2517. https://doi.org/10.1021/cm900783j.
[79] K. Matsunaga, K. Toyoura, First-principles analysis of oxide-ion conduction mechanism in lanthanum silicate, J Mater Chem 22 (2012) 7265–7273. https://doi.org/10.1039/C2JM16283K.

[80] K. Fukuda, T. Asaka, M. Okino, A. Berghout, E. Béchade, O. Masson, I. Julien, P. Thomas, Anisotropy of oxide-ion conduction in apatite-type lanthanum silicate, Solid State Ion 217 (2012) 40–45. https://doi.org/https://doi.org/10.1016/j.ssi.2012.04.018.
[81] K. Fukuda, T. Asaka, M. Oyabu, D. Urushihara, A. Berghout, E. Béchade, O. Masson, I. Julien, P. Thomas, Crystal structure and oxide-ion conductivity along c-axis of apatite-type lanthanum silicate with excess oxide ions, Chemistry of Materials 24 (2012) 4623–4631. https://doi.org/10.1021/cm3034643.
[82] S. Nakayama, Y. Higuchi, Y. Kondo, M. Sakamoto, Effects of cation- or oxide ion-defect on conductivities of apatite-type La–Ge–O system ceramics, Solid State Ion 170 (2004) 219–223. https://doi.org/https://doi.org/10.1016/j.ssi.2004.02.023.
[83] P.R. Slater, J.E.H. Sansom, J.R. Tolchard, Development of Apatite-Type Oxide, (2005) 373–384. https://doi.org/10.1002/tcr.20028.
[84] J.E.H. Sansom, P.R. Slater, Oxide ion conductivity in the mixed Si/Ge apatite-type phases La9.33Si6-xGexO26, Solid State Ion 167 (2004) 23–27. https://doi.org/https://doi.org/10.1016/j.ssi.2003.12.015.
[85] J.E.H. Sansom, A. Najib, P.R. Slater, Oxide ion conductivity in mixed Si/Ge-based apatite-type systems, Solid State Ion 175 (2004) 353–355. https://doi.org/https://doi.org/10.1016/j.ssi.2003.12.030.
[86] L. León-Reina, E.R. Losilla, M. Martínez-Lara, M.C. Martín-Sedeño, S. Bruque, P. Núñez, D. V Sheptyakov, M.A.G. Aranda, High oxide ion conductivity in Al-doped germanium oxyapatite, Chemistry of Materials 17 (2005) 596–600. https://doi.org/10.1021/cm048361r.
[87] L. León-Reina, J.M. Porras-Vázquez, E.R. Losilla, M.A.G. Aranda, Phase transition and mixed oxide-proton conductivity in germanium oxy-apatites, J Solid State Chem 180 (2007) 1250–1258. https://doi.org/https://doi.org/10.1016/j.jssc.2007.01.023.
[88] S.S. Pramana, W.T. Klooster, T.J. White, Framework `interstitial' oxygen in La10(GeO4)5(GeO5)O2 apatite electrolyte, Acta Crystallographica Section B 63 (2007) 597–602. https://doi.org/10.1107/S0108768107024317.
[89] J.E.H. Sansom, E. Kendrick, J.R. Tolchard, M.S. Islam, P.R. Slater, A comparison of the effect of rare earth vs Si site doping on the conductivities of apatite-type rare earth silicates, Journal of Solid State Electrochemistry 10 (2006) 562–568. https://doi.org/10.1007/s10008-006-0129-8.
[90] H. Zhang, F. Li, J. Jin, Q. Wang, Y. Sun, Synthesis and characterization of (Mg, Al)-doped apatite-type lanthanum germanate, Solid State Ion 179 (2008) 1024–1028. https://doi.org/https://doi.org/10.1016/j.ssi.2008.02.056.
[91] J.E.H. Sansom, P.A. Sermon, P.R. Slater, Synthesis and conductivities of the Ti doped apatite-type phases (La/Ba)10−x(Si/Ge)6−yTiyO26+z, Solid State Ion 176 (2005) 1765–1768. https://doi.org/https://doi.org/10.1016/j.ssi.2005.04.029.
[92] K. KOBAYASHI, Y. IGARASHI, N. SAITO, T. HIGUCHI, Y. SAKKA, T.S. SUZUKI, Stabilization of the high-temperature phase and total conductivity of yttrium-doped lanthanum germanate oxyapatite, Journal of the Ceramic Society of Japan 126 (2018) 91–98. https://doi.org/10.2109/jcersj2.17198.
[93] S.-F. Wang, Y.-F. Hsu, W.-J. Lin, K. Kobayashi, Transition metal-doped lanthanum germanate apatites as electrolyte materials of solid oxide fuel cells, Solid State Ion 247–248 (2013) 48–55. https://doi.org/https://doi.org/10.1016/j.ssi.2013.05.018.
[94] Q. Wang, Z. Huang, S. Chen, Enhancing conductivity of apatite-type lanthanum silicate by Li and Cu co-doping, J Phys Conf Ser 2351 (2022). https://doi.org/10.1088/1742-6596/2351/1/012010.
[95] H. Shang, Z. Huang, Enhanced electrical conductivity of P-doped apatite-type lanthanum silicate solid electrolytes and analysis of the mechanism, Ionics (Kiel) 29 (2023) 4013–4024. https://doi.org/10.1007/s11581-023-05154-0.
[96] J.R. Peet, M.S. Chambers, A. Piovano, M.R. Johnson, I.R. Evans, Dynamics in Bi(iii)-containing apatite-type oxide ion conductors: a combined computational and experimental study, J Mater Chem A Mater 6 (2018) 5129–5135. https://doi.org/10.1039/C8TA00546J.
[97] X. Yang, A.J. Fernández-Carrión, X. Kuang, Oxide Ion-Conducting Materials Containing Tetrahedral Moieties: Structures and Conduction Mechanisms, Chem Rev 123 (2023) 9356–9396. https://doi.org/10.1021/acs.chemrev.2c00913.
[98] A.C.R.V.V.I.L. Goulatis, Oxygen self-diffusion in apatites, (2012) 345–353.

https://doi.org/10.1007/s00706-011-0696-y.
[99] A. Najib, J.E.H. Sansom, J.R. Tolchard, P.R. Slater, M.S. Islam, Doping strategies to optimise the oxide ion conductivity in apatite-type ionic conductors, Dalton Transactions (2004) 3106–3109. https://doi.org/10.1039/B401273A.
[100] J.R. Tolchard, P.R. Slater, M.S. Islam, Insight into Doping Effects in Apatite Silicate Ionic Conductors, Adv Funct Mater 17 (2007) 2564–2571. https://doi.org/https://doi.org/10.1002/adfm.200600789.
[101] J. Li, Q. Cai, B.A. Horri, Synthesis and Densification of Mo/Mg Co-Doped Apatite-type Lanthanum Silicate Electrolytes with Enhanced Ionic Conductivity, Chemistry – A European Journal 29 (2023) e202300021. https://doi.org/https://doi.org/10.1002/chem.202300021.
[102] I. Perhaita, L.E. Muresan, G. Borodi, A. Popa, A. Nicoara, L.B. Tudoran, Studies on morpho-structure and ionic conductivity of apatite-type lanthanum silicate doped with transitional metal cations, J Electroceram (2024). https://doi.org/10.1007/s10832-024-00374-w.
[103] S. Fop, J.M.S. Skakle, A.C. McLaughlin, P.A. Connor, J.T.S. Irvine, R.I. Smith, E.J. Wildman, Oxide Ion Conductivity in the Hexagonal Perovskite Derivative Ba3MoNbO8.5, J Am Chem Soc 138 (2016) 16764–16769. https://doi.org/10.1021/jacs.6b10730.
[104] K.S. McCombie, E.J. Wildman, S. Fop, R.I. Smith, J.M.S. Skakle, A.C. McLaughlin, The crystal structure and electrical properties of the oxide ion conductor Ba3WNbO8.5, J Mater Chem A Mater 6 (2018) 5290–5295. https://doi.org/10.1039/c7ta08989a.
[105] S. Fop, K. McCombie, R.I. Smith, A.C. McLaughlin, Enhanced Oxygen Ion Conductivity and Mechanistic Understanding in Ba3Nb1- xV xMoO8.5, Chemistry of Materials 32 (2020) 4724–4733. https://doi.org/10.1021/acs.chemmater.0c01322.
[106] M. Yashima, T. Tsujiguchi, K. Fujii, E. Niwa, S. Nishioka, J.R. Hester, K. Maeda, Direct evidence for two-dimensional oxide-ion diffusion in the hexagonal perovskite-related oxide Ba3MoNbO8.5-δ, J Mater Chem A Mater 7 (2019) 13910–13916. https://doi.org/10.1039/c9ta03588e.
[107] Y. Yasui, T. Tsujiguchi, Y. Sakuda, J.R. Hester, M. Yashima, Oxide-Ion Occupational Disorder, Diffusion Path, and Conductivity in Hexagonal Perovskite Derivatives Ba3WNbO8.5 and Ba3MoNbO8.5, Journal of Physical Chemistry C 126 (2022) 2383–2393. https://doi.org/10.1021/acs.jpcc.1c09416.
[108] Y. Sakuda, T. Murakami, M. Avdeev, K. Fujii, Y. Yasui, J.R. Hester, M. Hagihala, Y. Ikeda, Y. Nambu, M. Yashima, Dimer-Mediated Cooperative Mechanism of Ultrafast-Ion Conduction in Hexagonal Perovskite-Related Oxides, Chemistry of Materials (2023). https://doi.org/10.1021/acs.chemmater.3c02378.
[109] S. Fop, K.S. McCombie, E.J. Wildman, J.M.S. Skakle, J.T.S. Irvine, P.A. Connor, C. Savaniu, C. Ritter, A.C. Mclaughlin, High oxide ion and proton conductivity in a disordered hexagonal perovskite, Nat Mater 19 (2020) 752–757. https://doi.org/10.1038/s41563-020-0629-4.
[110] Y. Sakuda, J.R. Hester, M. Yashima, Improved oxide-ion and lower proton conduction of hexagonal perovskite-related oxides based on $Ba_7Nb_4MoO_{20}$ by $Cr^{6+}$ doping, Journal of the Ceramic Society of Japan 130 (2022) 1–6.
[111] T. Murakami, T. Shibata, Y. Yasui, K. Fujii, J.R. Hester, M. Yashima, High Oxide-Ion Conductivity in a Hexagonal Perovskite-Related Oxide Ba7Ta3.7Mo1.3O20.15 with Cation Site Preference and Interstitial Oxide Ions, Small 18 (2022) 2106785. https://doi.org/https://doi.org/10.1002/smll.202106785.
[112] X. Turrillas, A.P. Sellars, B.C.H. Steele, Oxygen ion conductivity in selected ceramic oxide materials, Solid State Ion 28–30 (1988) 465–469. https://doi.org/https://doi.org/10.1016/S0167-2738(88)80084-5.
[113] E.J. Opila, H.L. Tuller, B.J. Wuensch, J. Maier, Oxygen Tracer Diffusion in La2-xSrxCuO4-y Single Crystals, Journal of the American Ceramic Society 76 (1993) 2363–2369. https://doi.org/https://doi.org/10.1111/j.1151-2916.1993.tb07778.x.
[114] J. Claus, G. Borchardt, S. Weber, J.-M. Hiver, S. Scherrer, Combination of EBSP measurements and SIMS to study crystallographic orientation dependence of diffusivities in a polycrystalline material: oxygen tracer diffusion in La2 − xSrxCuO4 ± δ, Materials Science and Engineering: B 38 (1996) 251–257.

https://doi.org/https://doi.org/10.1016/0921-5107(95)01446-2.
[115] V. V. Kharton, A. P. Viskup, E. N. Naumovich, F. M. B. Marques, Oxygen ion transport in La2NiO4-based ceramics, J Mater Chem 9 (1999) 2623–2629. https://doi.org/10.1039/A903276B.
[116] B. Dabrowski, J.D. Jorgensen, D.G. Hinks, S. Pei, D.R. Richards, H.B. Vanfleet, D.L. Decker, La2CuO4+δ and La2NiO4+δ: Phase separation resulting from excess oxygen defects., Physica C: Superconductivity and Its Applications 162–164 (1989) 99–100. https://doi.org/https://doi.org/10.1016/0921-4534(89)90936-2.
[117] D.E. Rice, D.J. Buttrey, An X-Ray Diffraction Study of the Oxygen Content Phase Diagram of La2NiO4+δ, J Solid State Chem 105 (1993) 197–210. https://doi.org/https://doi.org/10.1006/jssc.1993.1208.
[118] H. Tamura, A. Hayashi, Y. Ueda, Phase diagram of La2NiO4+δ ($0 \leq \delta \leq 0.18$): I. Phase at room temperature and phases transition above $\delta = 0.15$, Physica C Supercond 216 (1993) 83–88. https://doi.org/https://doi.org/10.1016/0921-4534(93)90636-5.
[119] M.T. Fernández-Díaz, J.L. Martínez, J. Rodríguez-Carvajal, High-temperature phase transformation of oxidized R2NiO4+δ(R=La, Pr and Nd) under vacuum, Solid State Ion 63–65 (1993) 902–906. https://doi.org/https://doi.org/10.1016/0167-2738(93)90213-M.
[120] S.J. Skinner, J.A. Kilner, Oxygen diffusion and surface exchange in La2−xSrxNiO4+δ, Solid State Ion 135 (2000) 709–712. https://doi.org/https://doi.org/10.1016/S0167-2738(00)00388-X.
[121] X. Li, N.A. Benedek, Enhancement of ionic transport in complex oxides through soft lattice modes and epitaxial strain, Chemistry of Materials 27 (2015) 2647–2652. https://doi.org/10.1021/acs.chemmater.5b00445.
[122] S. Xu, R. Jacobs, D. Morgan, Factors Controlling Oxygen Interstitial Diffusion in the Ruddlesden–Popper Oxide La2–xSrxNiO4+δ, Chemistry of Materials 30 (2018) 7166–7177. https://doi.org/10.1021/acs.chemmater.8b03146.
[123] M. Yashima, M. Enoki, T. Wakita, R. Ali, Y. Matsushita, F. Izumi, T. Ishihara, Structural disorder and diffusional pathway of oxide ions in a doped Pr2NiO4-based mixed conductor., J Am Chem Soc 130 (2008) 2762–2763. https://doi.org/10.1021/ja711478h.
[124] A. Chroneos, D. Parfitt, J.A. Kilner, R.W. Grimes, Anisotropic oxygen diffusion in tetragonal La2NiO 4+δ: Molecular dynamics calculations, J Mater Chem 20 (2010) 266–270. https://doi.org/10.1039/b917118e.
[125] A. Kushima, D. Parfitt, A. Chroneos, B. Yildiz, J.A. Kilner, R.W. Grimes, Interstitialcy diffusion of oxygen in tetragonal La2CoO4+δ, Physical Chemistry Chemical Physics 13 (2011) 2242–2249. https://doi.org/10.1039/C0CP01603A.
[126] A. Perrichon, A. Piovano, M. Boehm, M. Zbiri, M. Johnson, H. Schober, M. Ceretti, W. Paulus, Lattice Dynamics Modified by Excess Oxygen in Nd2NiO4+δ: Triggering Low-Temperature Oxygen Diff usion, (2015).
[127] D. Lee, H.N. Lee, Controlling oxygen mobility in ruddlesden-popper oxides, Materials 10 (2017) 1–22. https://doi.org/10.3390/ma10040368.
[128] M. Burriel, G. Garcia, J. Santiso, J.A. Kilner, R.J. Chater, S.J. Skinner, Anisotropic oxygen diffusion properties in epitaxial thin films of La2NiO4+δ, J Mater Chem 18 (2008) 416–422. https://doi.org/10.1039/b711341b.
[129] J. Bassat, O. Wahyudi, M. Ceretti, P. Veber, I. Weill, A. Villesuzanne, J. Grenier, W. Paulus, J.A. Kilner, Anisotropic Oxygen Diffusion Properties in Pr2NiO4+δ and Nd2NiO4+δ Single Crystals, The Journal of Physical Chemistry C (2013).
[130] D. Parfitt, A. Chroneos, J.A. Kilner, R.W. Grimes, Molecular dynamics study of oxygen diffusion in Pr2NiO4+δ, Physical Chemistry Chemical Physics 12 (2010) 6834–6836. https://doi.org/10.1039/c001809k.
[131] M. Rozumek, P. Majewski, H. Schluckwerder, F. Aldinger, K. Künstler, G. Tomandl, Electrical Conduction Behavior of La1+Sr1−Ga3O7–δ Melilite-Type Ceramics, Journal of the American Ceramic Society 87 (2004) 1795–1798. https://doi.org/https://doi.org/10.1111/j.1551-2916.2004.01795.x.
[132] J. Schuett, T.K. Schultze, S. Grieshammer, Oxygen Ion Migration and Conductivity in LaSrGa3O7Melilites from First Principles, Chemistry of Materials 32 (2020) 4442–4450. https://doi.org/10.1021/acs.chemmater.9b04599.
[133] F. Wei, H. Gasparyan, P.J. Keenan, M. Gutmann, Y. Fang, T. Baikie, J.B. Claridge, P.R.

Slater, C.L. Kloc, T.J. White, Anisotropic oxide ion conduction in melilite intermediate temperature electrolytes, J Mater Chem A Mater 3 (2015) 3091–3096. https://doi.org/10.1039/c4ta05132g.
[134] B. Liu, D. Ding, Z. Liu, F. Chen, C. Xia, Synthesis and electrical conductivity of various melilite-type electrolytes Ln1+xSr1−xGa3O7+x/2, Solid State Ion 191 (2011) 68–72. https://doi.org/https://doi.org/10.1016/j.ssi.2011.04.005.
[135] L. Zhou, J. Xu, M. Allix, X. Kuang, Development of Melilite-Type Oxide Ion Conductors, Chemical Record (2020) 1–13. https://doi.org/10.1002/tcr.202000069.
[136] J. Xu, J. Wang, X. Tang, X. Kuang, M.J. Rosseinsky, La1+xBa1-xGa3O7+0.5x Oxide Ion Conductor: Cationic Size Effect on the Interstitial Oxide Ion Conductivity in Gallate Melilites, Inorg Chem 56 (2017) 6897–68905. https://doi.org/10.1021/acs.inorgchem.7b00295.
[137] Y. Li, H. Yi, J. Xu, X. Kuang, High oxide ion conductivity in the Bi3+ doped melilite LaSrGa3O7, J Alloys Compd 740 (2018) 143–147. https://doi.org/10.1016/j.jallcom.2017.12.363.
[138] F.H. Danmo, A. Westermoen, K. Marshall, D. Stoian, T. Grande, J. Glaum, S.M. Selbach, High-Entropy Hexagonal Manganites for Fast Oxygen Absorption and Release, Chemistry of Materials 36 (2024) 2711–2720. https://doi.org/10.1021/acs.chemmater.3c02702.
[139] F.H. Danmo, I.E. Nylund, A. Westermoen, K.P. Marshall, D. Stoian, T. Grande, J. Glaum, S.M. Selbach, Oxidation Kinetics of Nanocrystalline Hexagonal RMn1-xTixO3 (R = Ho, Dy), ACS Appl Mater Interfaces 15 (2023) 42439–42448. https://doi.org/10.1021/acsami.3c06020.
[140] S. Remsen, B. Dabrowski, Synthesis and oxygen storage capacities of hexagonal Dy1-xY xMnO3+δ, Chemistry of Materials 23 (2011) 3818–3827. https://doi.org/10.1021/cm2006956.
[141] J. JEEVARATNAM, L.S.D. GLASSER, F.P. GLASSER, Structure of Calcium Aluminate, 12CaO.7AL2O3, Nature 194 (1962) 764–765. https://doi.org/10.1038/194764b0.
[142] Y. Lv, Y. Sun, J. Xu, X. Xu, A.J. Fernández-Carrión, T. Wei, H. Yi, X. Kuang, Phase Evolution, Electrical Properties, and Conduction Mechanism of Ca12Al14- xGa xO33(0 ≤ x ≤ 14) Ceramics Synthesized by a Glass Crystallization Method, Inorg Chem 60 (2021) 2446–2456. https://doi.org/10.1021/acs.inorgchem.0c03344.
[143] H. Hosono, K. Hayashi, K. Kajihara, P. V. Sushko, A.L. Shluger, Oxygen ion conduction in 12CaO·7Al2O3: O2- conduction mechanism and possibility of O- fast conduction, Solid State Ion 180 (2009) 550–555. https://doi.org/10.1016/j.ssi.2008.10.015.
[144] M. Teusner, R.A. De Souza, H. Krause, S.G. Ebbinghaus, B. Belghoul, M. Martin, Oxygen Diffusion in Mayenite, Journal of Physical Chemistry C 119 (2015) 9721–9727. https://doi.org/10.1021/jp512863u.
[145] H. Yi, Y. Lv, V. Mattick, J. Xu, Cobalt-doped Ca12Al14O33 mayenite oxide ion conductors: phases, defects, and electrical properties, Ionics (Kiel) 25 (2019) 5105–5115. https://doi.org/10.1007/s11581-019-03088-0.
[146] H. Yi, Y. Lv, Y. Wang, X. Fang, V. Mattick, J. Xu, Ga-doped Ca 12 Al 14 O 33 mayenite oxide ion conductors: synthesis, defects, and electrical properties, RSC Adv 9 (2019) 3809–3815. https://doi.org/10.1039/c8ra08254e.
[147] J. Janek, D.K. Lee, Defect chemistry of the mixed conducting cage compound Ca 12Al14O33, Journal of the Korean Ceramic Society 47 (2010) 99–105. https://doi.org/10.4191/KCERS.2010.47.2.099.
[148] H. Seh, H.L. Tuller, Defects and transport in langasite I: Acceptor-doped (La3Ga 5SiO14), J Electroceram 16 (2006) 115–125. https://doi.org/10.1007/s10832-006-4081-x.
[149] Z. Luo, X. Li, X. Wang, S. Deng, L. He, K. Lin, Q. Li, X. Xing, X. Kuang, The Langasite Family for the Development of Oxygen-Vacancy-Mediated Oxide Ion Conductors, Chemistry of Materials 36 (2024) 2835–2845. https://doi.org/10.1021/acs.chemmater.3c03148.
[150] S. Kajita, N. Ohba, A. Suzumura, S. Tajima, R. Asahi, Discovery of superionic conductors by ensemble-scope descriptor, NPG Asia Mater 12 (2020) 1–8. https://doi.org/10.1038/s41427-020-0211-1.
[151] S. Tajima, N. Ohba, A. Suzumura, S. Kajita, Synthesis and Ion-Transport Properties of EuKGe2O6-, Ca3Fe2Ge3O12-, and BaCu2Ge2O7-Type Oxide-Ion Conductors, Inorg

Chem 60 (2021) 17019–17032. https://doi.org/10.1021/acs.inorgchem.1c02143.
[152] Y. Yasui, E. Niwa, M. Matsui, K. Fujii, M. Yashima, Discovery of a Rare-Earth-Free Oxide-Ion Conductor Ca3Ga4O9 by Screening through Bond Valence-Based Energy Calculations, Synthesis, and Characterization of Structural and Transport Properties, Inorg Chem 58 (2019) 9460–9468. https://doi.org/10.1021/acs.inorgchem.9b01300.
[153] Y. Yasui, K. Jojima, K. Fujii, K. Mori, M. Yashima, High Oxide-Ion Conduction in Rb-Containing Oxides, Chemistry of Materials (2025). https://doi.org/10.1021/acs.chemmater.4c03148.